\documentclass[final,authoryear,12pt]{elsarticle}
\usepackage{graphicx}
\usepackage{amsthm}
\usepackage[fleqn]{amsmath}
\usepackage{subfigure}
\usepackage{epstopdf}
\usepackage{multirow}
\usepackage{booktabs} 
\usepackage{slashbox}
\usepackage[draft,bookmarks]{hyperref}
\hypersetup{colorlinks=true,linkcolor=black,urlcolor=black,citecolor=black,bookmarks=true}
\usepackage{geometry}                		
\usepackage{graphicx}				
\usepackage{lscape}								
\usepackage{amssymb}
\usepackage[usenames]{xcolor} 
 
\biboptions{comma,round}

\journal{Working Paper}

\begin{document}

\begin{frontmatter}

\title{Are benefits from oil -- stocks diversification gone? New evidence from a dynamic copula and high frequency data\tnoteref{label1}}

\author[ies,utia]{Krenar Avdulaj}
 \author[ies,utia]{Jozef Barunik\corref{cor2}} \ead{barunik@utia.cas.cz}
 \address[ies]{Institute of Information Theory and Automation, Academy of Sciences of the Czech Republic, Czech Republic}
\address[utia]{Institute of Economic Studies, Charles University, Prague, Czech Republic}

\tnotetext[label1]{Support from the Czech Science Foundation under the 13-32263S and 13-24313S projects and the Grant Agency of the Charles University under the 852013 project is gratefully acknowledged.}

\begin{abstract}
Oil is perceived as a good diversification tool for stock markets. To fully understand this potential, we propose a new empirical methodology that combines generalized autoregressive score copula functions with high frequency data and allows us to capture and forecast the conditional time-varying joint distribution of the oil -- stocks pair accurately. Our realized GARCH with time-varying copula yields statistically better forecasts of the dependence and quantiles of the distribution relative to competing models. Employing a recently proposed conditional diversification benefits measure that considers higher-order moments and nonlinear dependence from tail events, we document decreasing benefits from diversification over the past ten years. The diversification benefits implied by our empirical model are, moreover, strongly varied over time. These findings have important implications for asset allocation, as the benefits of including oil in stock portfolios may not be as large as perceived.
\end{abstract}

\end{frontmatter}

\section{Introduction}

The risk reduction benefit from diversification has been a major subject in the finance literature for decades. The number of studies exploring the role of oil prices in equity returns remains limited, with no consensus regarding the nature and number of factors that play a role in determining equity returns. Despite the rather scarce literature, the idea of utilizing crude oil as a diversification tool for financial assets attracted number of publications 
\citep{gorton2006facts,buyuksahin2008commodities,fratzscher2014oil}, which conclude that oil is a nearly perfect diversification tool for stocks due to their null, or even negative, correlation. This feature is also reflected in investorsÕ demand for products to diversify risk. For example Morgan Stanley offers a product composed of oil and S\&P 500 prices with equal weights. After the recent financial turmoil in 2008, the literature began to document a possible reduction of these benefits due to rising dependence and financialization of commodities \citep{tang2012index,buyukcsahin2014speculators}. However, a majority of empirical studies use linear (time-varying) correlations to measure benefits, ignoring the possible non-linear dependence of tail events. Hence, it is natural to ask the question posed in the title of this paper: ``Are the benefits from oil-stocks diversification gone?" To answer this question, a proper measure of the benefits, which will account for the following two key issues, must be employed. 

First, the time-varying nature of the correlations must be addressed properly. The recent turmoil in financial markets, which began in September 2008, further strengthened the focus on models that are able to capture dynamic dependencies in data.\footnote{\cite{miller2009crude} analyze the long-run relationship between oil and international stock markets utilizing cointegration techniques, and they find that stock markets responded negatively to increases in oil prices in the long run before 1999, but after this point, the relationship collapsed. This finding is in line with a number of studies reporting the negative influence of rising oil prices on stock markets \citep{sadorsky1999oil,ciner2001energy,nandha2008does,o2008role,park2008oil,chen2010higher}. Generally, these results are consistent with economic theory, as rising oil prices are expected to have an adverse effect on real output and, hence, an adverse effect on corporate profits if oil is employed as a key input. This phenomenon suggests that oil could be an important factor for equity returns. In a large study of the relationship among oil prices, exchange rates and emerging stock markets, \cite{basher2012oil} confirm previous research by finding that a positive shock in oil prices tends to depress stock markets and U.S. dollar exchange rates in the short run. \cite{wu2012economic} further study the depreciation in the U.S. dollar causing an increase in crude oil prices. \cite{geman2008wti}  study the maturity effect in the choice of oil futures with respect to diversification benefits and find that futures with more distant maturities are more negatively correlated with the S\&P 500. More recently, new evidence from data during and after the 2008 financial crisis has emerged. \cite{mollick2012us} find that, while stock returns were negatively affected by oil prices prior to the crisis, this relationship became positive after 2009. The authors interpret this reversal as stocks positively responding to expectations of recovery. \cite{hammoudeh2013time} study the dynamics of oil and Central and Eastern European (CEE) stock markets, and they find a positive time-varying relationship utilizing recent data. \cite{wen2012measuring} finds contagion between energy and stock markets that arose during the 2008 financial crisis. Finally, \cite{broadstock2012oil,sadorsky2012correlations} document price and volatility spillovers in oil and stocks.} 

Second, linear correlation may not be a satisfactory measure of dependence, as it does not account for dependence between tail events. Correlation asymmetries and changes in correlation due to business cycle conditions are crucial, as these dependences will impact to a large degree the benefits from diversification.

In this paper, we provide a comprehensive empirical study of the dynamics in dependence and tail dependence and translate the results directly to the diversification benefits. We offer two contributions. First, we propose a flexible empirical model for oil-stock dependence, which couples time-varying copula models with high frequency data. Employing the increasingly popular Generalized Autoregressive Score (GAS) framework \citep{Creal2013}, we model the joint distribution of oil-stock returns utilizing time-varying copula functions and capture nonlinear dependence. This is an important step to take, as the dynamics of correlations may depart from the one imposed by the assumption of multivariate normality used in many different approaches. In addition, we employ the recently developed realized GARCH model \citep{Hansen2012}, which uses high frequency-based measures of volatility to better capture the volatility process in the margins of the oil-stocks return distribution. Our newly proposed empirical model, the realized GARCH time-varying GAS copula, is thus a very flexible approach. Moreover, we employ a semiparametric alternative to modeling strategy, which combines a nonparametric estimate of a margins distribution and parametric copula function. While this approach is empirically attractive, it is not often employed in the literature.

Second, we study conditional diversification benefits, which are implied by our model using the appealing framework of \cite{christoffersen2012potential}. This framework considers higher-order moments and non-linear dependence, which is an important step to take, as diversification benefits implied by simple linear correlations will likely be under- or over-estimated, depending on the degree of dependence coming from the tails. In addition, we evaluate our empirical model utilizing Value-at-Risk forecasts.

We demonstrate that our newly proposed empirical model is able to capture and forecast the time-varying dynamics in a joint distribution accurately. We revisit the oil-stock relationship utilizing the large span of data covering the periods for which the literature finds a negative relationship, as well as the recent stock market turmoil. The main finding is that we document decreasing benefits from the usage of oil as a diversification tool for stocks until the year 2011, which can be attributed to the changing expectations of investors after the recent market turmoil. After the year 2011, diversification benefits started to increase slowly and showed promise to investors who wish to use oil as a hedge for their stock portfolio. Moreover, in an empirical section, we test the economic significance of the model and demonstrate that our method yields more accurate quantile forecasts, which are central to risk management due to the popular Value-at-Risk (VaR) measure. 

The work is organized as follows. The second section introduces our empirical model based on a dynamic copula realized GARCH modeling framework in detail. The third section introduces the data, and the fourth section offers empirical results documenting the time-varying nature of dependence between oil and stocks and good out-of-sample performance of models. The fifth section then elaborates on the economic implications of our modeling strategy, employing quantile forecasts, quantifying the risk of an equally weighted portfolio composed of oil and stocks and, finally, documents the time variation in the benefits of utilizing oil as a diversification tool for stocks. The last section concludes. 

\section{Dynamic copula realized GARCH modeling framework}
Our modeling strategy utilizes high frequency data to capture the dependence in the margins and recently proposed dynamic copulas to model the dynamic dependence. The final model is thus able to describe the conditional time-varying joint distribution of oil and stocks, which will be very useful in the economic application.

The methodology employed in this work is based on \citeauthor{Sklar1959}'s (1959) theorem extended to conditional distributions by \cite{Patton2006}.  \citeauthor{Sklar1959}'s extended theorem allows us to decompose a conditional joint distribution into marginal distributions and a copula. Consider the bivariate stochastic process $\{\mathbf{X}_t\}_{t=1}^T$ with $\mathbf{X_t} = (X_{1t},X_{2t})'$, which has a conditional joint distribution $\mathbf{F_t}$ and conditional marginal distributions $F_{1t}$ and $F_{2t}$. Then
\begin{equation}
\mathbf{X_t}|\mathcal{F}_{t-1} \sim \mathbf{F_t} = \mathbf{C_t}\left(F_{1t},F_{2t}\right),
\end{equation}
where $\mathbf{C_t}$ is the time-varying conditional copula of $\mathbf{X_t}$ containing all information about the dependence between $X_{1t}$ and $X_{2t}$, and $\mathcal{F}_{t-1}$ the information set. Due to \citeauthor{Sklar1959}'s theorem, we are thus able to construct a dynamic joint distribution $\mathbf{F_t}$ by linking together any two marginal distributions $F_{1t}$ and $F_{2t}$ with any copula function providing very flexible approach for modeling joint dynamic distributions.\footnote{Please note that the information set for the margins and the copula conditional density is the same.}

\subsection{Time-varying conditional marginal distribution with realized measures}

The first step in building an empirical model based on copulas is to find a proper model for the marginal distributions. As the most pronounced dependence that can be found in the returns time series is the one in variance, the vast majority of the literature utilizes the conventional generalized autoregressive conditional heteroscedasticity (GARCH) approach of \cite{Boll86} in this step. 

With the increasing availability of high frequency data, the literature moved to a different concept of volatility modeling called realized volatility. This very simple and intuitive approach for computing daily volatility employing the sum of squared high-frequency returns was formalized by \cite{abdl2003} and \cite{barndorff2004b}. While realized volatility can be measured simply from the high frequency data, one must specify a correct model to be able to use this parameter for forecasting. In past years, researchers found ways to include a realized volatility measure to assist GARCH-type parametric models in capturing volatility. 

As noted previously, the key object of interest in financial econometrics, the conditional variance of returns, $h_{it}=var(X_{it}|\mathcal{F}_{t-1})$, is usually modeled by GARCH. While in a standard GARCH(1,1) model the conditional variance, $h_{it}$, is dependent on its past values $h_{it-1}$ and the values of $X_{it-1}^2$, \cite{Hansen2012} propose to utilize a realized volatility measure and make $h_{it}$ dependent on the realized variance. The authors propose a so-called measurement equation that ties the realized measure to latent volatility. The general framework of realized GARCH($p$,$q$) models is well connected to the existing literature in \cite{Hansen2012}. Here, we restrict ourselves to the simple log-linear specification of the realized GARCH(1,1). While it is important to model the conditional time-varying mean $E(X_{it}|\mathcal{F}_{t-1})$, we also include the standard AR model in the final modeling strategy. As we will discuss later, the autoregressive term of order no larger than two is appropriate for the oil and stocks data in the study; thus, we restrict ourselves to specifying the AR(2) with log-linear realized GARCH(1,1) model as in \cite{Hansen2012}  
\begin{align}
\label{realizedgarch}
X_{it}&=\mu_i + \alpha_1X_{it-1} + \alpha_2X_{it-2} +  \sqrt{h_{it}}z_{it}, \hspace{3cm} \mbox{for  }  i=1,2 \\
\log h_{it}&=\omega_i + \beta_i \log h_{it-1} + \gamma_i \log RV_{it-1}, \\
\log RV_{it}&=\psi_i + \phi_i \log h_{it} + \tau_i(z_{it}) + u_{it},
\end{align}

where $\mu_i$ is the constant mean, $h_{it}$ conditional variance, which is latent, $RV_{it}$ realized volatility measure, $u_{it} \sim N(0,\sigma^2_{iu})$, and $\tau_i(z_{it})=\tau_{i1} z_{it} + \tau_{i2} (z_{it}^2-1)$ leverage function. For the $RV_{it}$, we employ the high frequency data and compute it as a sum of squared intraday returns \citep{abdl2003,barndorff2004b}. We will provide more detail on how we compute the realized volatility measure in the empirical section. \cite{Hansen2012} suggests estimating the parameters utilizing a quasi-maximum likelihood estimator (QMLE), which is very similar to the standard GARCH. While we have realized measures in the estimation yielding additional measurement error $u_{it}$, we need to factorize the joint conditional density\footnote{Please note that information set $\mathcal{F}_{t-1}$ contains the lagged values of $RV_{it}$ as well.} $f(X_{it},RV_{it}|\mathcal{F}_{t-1})=f(X_{it}|\mathcal{F}_{t-1})f(RV_{it}|X_{it},\mathcal{F}_{t-1})$ which results in a sum after logarithmic transform and thus, is readily available for finding parameters. In our model, we allow the innovations $z_{it}$ to follow skewed-\emph{t} distribution of \cite{Hansen1994}, having two shape parameters, a skewness parameter $\lambda\in(-1,1)$ controlling the degree of asymmetry, and a degree of freedom parameter $\nu\in(2,\infty]$ controlling the thickness of tails. When $\lambda=0$, the distribution becomes the standard Student's $t$ distribution,  when $\nu \rightarrow\infty$, it becomes skewed Normal distribution, while for  $\nu \rightarrow\infty$ and $\lambda=0$, it becomes $N(0,1)$. Thus, the choice of the skewed-\emph{t} distribution gives us flexibility to capture the potential measurement errors from realized volatility and hence, possible departures from the normality of residuals.

Thus, after the time-varying dependence in the mean and volatility is modeled, we are left with residuals
\begin{align}
 \hat{z}_{it}&=\frac{X_{it}-\hat{\mu}_i-\hat{\alpha}_1X_{it-1} - \hat{\alpha}_2X_{it-2}}{ \sqrt{\hat{h}_{it}}} \\
 \hat{z}_{it}|\mathcal{F}_{t-1}&  \sim F_i(0,1), \hspace{3cm} \hbox{for } i=1,2.
\end{align}
which have a constant conditional distribution with zero mean and variance one.  Then, the conditional copula of $\mathbf{X_t}|\mathcal{F}_{t-1}$ is equal to the conditional distribution of $\mathbf{U_t}|\mathcal{F}_{t-1}$:
\begin{equation}
\mathbf{U_t}|\mathcal{F}_{t-1} \sim \mathbf{C_t(\gamma_0)},
\end{equation}
with $\gamma$ being copula parameters, and $\mathbf{U_t}=[U_{1t},U_{2t}]'$ conditional probability integral transform 
\begin{equation}
U_{it} = F_{i} \left(\hat{z}_{it};\phi_{i,0} \right), \hspace{3cm} \hbox{for } i=1,2.
\end{equation}

\subsection{Dynamic copulas: A ``GAS" dynamics in parameters}

After finding a model for the marginal distribution, we proceed to the copula functions. An important feature that is required for our work is the specification that parameters are allowed vary over time. Recently, \cite{hafner2012,manner2011} proposed a stochastic copula model that allows the parameters to evolve as a latent time series. Another possibility is offered by ARCH-type models for volatility \citep{Engle2002} and related models for copulas \citep{Patton2006,Creal2013}, which allow the parameters to be some function of lagged observables. An advantage of the second approach is that it avoids the need to ``integrate out" the innovation terms driving the latent time series processes. 

For our empirical model, we adopt the generalized autoregressive score (GAS) model of \cite{Creal2013}, which specifies the time-varying copula parameter ($\delta_t$) as a function of the lagged copula parameter and a forcing variable that is related to the standardized score of the copula log-likelihood\footnote{ \cite{harvey2013,harvey2012} propose a similar method for modeling time-varying parameters, which they call a dynamic conditional score model.}. 
Consider a copula with time-varying parameters:
\begin{equation}
\mathbf{U_t}|\mathcal{F}_{t-1} \sim \mathbf{C_t(\delta_t(\gamma))}.
\end{equation}

Often, a copula parameter is required to fall within a specific range, e.g., the correlation for Normal or $t$ copula is required to fall in between values of -1 and 1. To ensure this, \cite{Creal2013} suggest transforming the copula parameter by an increasing invertible function\footnote{For example, for the Normal and $t$ copula, the transformation is $(1-e^{-\kappa_t})/(1+e^{-\kappa_t})$. Concrete functions for other copulas employed in our work are given in Appendix A, which introduces copula functions.} (e.g., logarithmic, logistic, etc.) to the parameter:
\begin{equation}
\kappa_t=h(\delta_t) \Longleftrightarrow  \delta_t= h^{-1}(\kappa_t) \label{eq:invertft}
\end{equation}
For a copula with transformed time-varying parameter $\kappa_t$, a GAS(1,1) model is specified as
\begin{align}
 \kappa_{t+1}&= w + \beta \kappa_{t}+ \alpha I^{-1/2}_t \mathbf{s}_{t} \\
\mathbf{s}_{t} &\equiv\displaystyle \frac{ \partial \log \mathbf{c} (\mathbf{u_{t}};\delta_{t})}{\partial \delta_{t}} \\
I_t &\equiv E_{t-1}[\mathbf{s}_{t}\mathbf{s}_{t}']=I(\delta_t).
\end{align}
While this specification for the time-varying parameters is arbitrary, \cite{Creal2013} motivates it in a way that the model nests a variety of popular approaches from conditional variance models to trade durations and counts models. Additionally, the recursion is similar to numerical optimization algorithms such as the Gauss-Newton algorithm. In comparison to the approach of \cite{Patton2006}, the GAS specification implies more sensitivity to correlation shocks. Hence, reactions from returns of an opposite sign in the situation of a positive correlation estimate will be captured. For more details, and empirical comparison of the two approached, see Section 3.1. in \cite{Creal2013}.

Until now, we have focused attention on the specification of the dynamics of the models. What remains to be specified is the shape of the copula. In our modeling strategy, we will compare several of the most often utilized shapes of copula functions, while the rest of the model will be fixed. For the dynamic parameter models, we will employ the rotated Gumbel, Normal and Student's $t$ functional forms described briefly in the \ref{appA}. In our empirical application, we also employ constant copula functions as a benchmark. These are described in the \ref{appA} as well.

\subsection{Estimation strategy}
	
The final dynamic copula realized GARCH model defines a dynamic parametric model for the joint distribution. The joint likelihood is
	\begin{align}
	\mathcal{L}(\theta)   \equiv \sum_{t=1}^T\log \mathbf{f}_t(\mathbf{X_t};\theta) =& \sum_{t=1}^T \log f_{1t}(X_{1t};\theta_1)+\sum_{t=1}^T \log f_{2t}(X_{2t};\theta_2) \\
	&+\sum_{t=1}^T \log \mathbf{c}_t(F_{1t}(X_{1t};\theta_1),F_{2t}(X_{2t};\theta_2);\theta_c),
	\end{align}
	where $\theta=(\phi',\gamma')'$ is vector of all parameters to be estimated, including parameters of the marginal distributions $\phi$ and parameters of the copula, $\gamma$. The parameters are estimated utilizing a two-step estimation procedure, generally known as multi-stage maximum likelihood (MSML) estimation, first estimating the marginal distributions and then estimating the copula model conditioning on the estimated marginal distribution parameters. While this greatly simplifies the estimation, inference on the resulting copula parameter estimates is more difficult than usual as the estimation error from the marginal distribution must be considered. As a result, MSMLE is asymptotically less efficient than one-stage MLE; however, as discussed by \cite{patton2006b}, this loss is small in many cases. Moreover, the bootstrap methodology can be utilized, as discussed in following sections. 
	
\subsubsection{Semiparametric models}
	
	One of the appealing alternatives to a fully parametric model is to estimate univariate distribution non-parametrically, for example, by utilizing the empirical distribution function. Combination of a nonparametric model for marginal distribution and parametric model for the copula results in a semiparametric copula model, which we use for comparison to its fully parametric counterpart. In our modeling strategy, we concentrate on a full parametric model combining fully parametric marginal distribution $F_i$ with a copula function, while the theory is developed for the inference. Still, a nonparametric distribution $F_i$ has great empirical appeal; thus, we utilize it for comparison and rely on bootstrap-based inference for parameter estimates, as discussed later in the text. Forecasts based on a semiparametric estimation where nonparametric marginal distribution is combined with parametric copula function are not common in economic literature, thus, it is interesting to compare it in our modeling strategy. For the margins of the semi-parametric models, we employ the non-parametric empirical distribution $F_i$ introduced by \cite{Genest:1995fk}\footnote{The asymptotic properties of this estimator can be found in \cite{Chen:2006kx}.}, which consists of modeling the marginal distributions by the (rescaled) empirical distribution.
\begin{equation}
\hat{F}_i(z)= \frac{1}{T+1}  \sum_{t=1}^T \mathbf{1} \{\hat{z}_{it} \leq z\}
\end{equation}
In this case, the parameter estimation is conducted by maximizing likelihood
\begin{equation}
	\mathcal{L}(\gamma)   \equiv \sum_{t=1}^T \log \mathbf{c}_t(\hat{U}_{1t},\hat{U}_{2t};\gamma),
	\end{equation}
Again, the inference of parameters is more difficult than usual. We discuss the inference in the following section. 
	
	\subsection{Inference for parameter estimates}
	
	For the statistical inference of parameters, we utilize the bootstrapping methodology as suggested by \cite{Patton2006}. More specifically, for constant parametric copulas, we employ the stationary bootstrap of \cite{Politis1994}, while for the constant semi-parametric \emph{i.i.d.} bootstrapping. The use of these bootstrapping methods is justified by the work of \cite{Goncalves2004,Chen2006} and \cite{Remillard2010}. The algorithm utilized to obtain the statistical inference for parametric model (both constant and time-varying) follows these steps:
\begin{enumerate}[i)]
\item Use a block bootstrap to generate a bootstrap sample of the data of length T.
\item Estimate the model using the same multi-stage approach as applied for the real data.
\item Repeat steps (i)-(ii) S times\footnote{We use S=100 due to the high computational power needed for time-varying \emph{t} copula and because larger S in fact does not substantially improve the results (these ``testing" results with S=1000 are available upon request from authors).}.
\item Use the $\alpha/2$ and $1-\alpha/2$ quantiles of the distribution of estimated parameters to obtain a $1-\alpha$ confidence interval for these parameters. \end{enumerate} 

For the constant semi-parametric copulas the algorithm follows:
\begin{enumerate}[i)]
\item Use an \emph{i.i.d.} bootstrap to generate a bootstrap sample of the estimated standardized residuals of length T.
\item Transform each time series of bootstrap data using its empirical distribution function.
\item Estimate the copula model on the transformed data.
\item Repeat steps (i)-(iii) S times.
\item Use the $\alpha/2$ and $1-\alpha/2$ quantiles of the distribution of estimated parameters to obtain a $1-\alpha$ confidence interval for these parameters.
\end{enumerate}

When we consider semi-parametric time-varying copulas, we cannot utilize the \emph{iid} bootstrap because the true standardized residuals are not jointly \emph{iid}. Inference methods for these models are not yet available. However, \cite{Patton2012} suggests employing the block bootstrap technique (\emph{e.g.} stationary bootstrap of \cite{Politis1994}), stressing the need for formal justification. 

\subsection{Goodness-of-fit and copula selection}
A crucial issue in empirical copula applications is related to the goodness-of-fit. While copula models allow great flexibility, it is crucial to find the model that is well specified for the data as more harm then help can be done when one relies on a misspecified model.
 \cite{Genest2009} make a review on available goodness-of-fit tests for copulas. Two tests that are widely used for goodness-of-fit tests of copula models and that we utilize are the standard Kolmogorov-Smirnov (KS) and Cramer von-Mises (CvM) tests. These approaches work only for constant copula models. When dealing with time-varying copulas we should modify the testing procedure. Thus, we utilize the fitted copula model to obtain the Rosenblatt transform of the data, which is a multivariate version of the probability integral transformation. In the multivariate version, these tests then measure the distance between the empirical copula estimated on Rosenblatt's transformed data denoted by $\mathbf{\hat C}_T$ and the independence copula denoted by  $\mathbf{C}_{\bot}$.

Rosenblatt's probability integral transform of a copula $\mathbf{C}$ is the mapping $\mathcal{R}: (0,1)^n  \rightarrow (0,1)^n$. To every $\mathbf{U}_t=(U_{1t},\ldots,U_{nt}) \in (0,1)^n$, this mapping assigns another vector $\mathcal{R}(\mathbf{U}_t)=(V_{1t},\ldots,V_{nt})$ with $V_{1t}=U_{1t}$ and for each $i \in \{2, \ldots, n\}$,
\begin{equation}
V_{it}= \displaystyle \frac{\partial^{i-1} C(U_{1t},\ldots,U_{it},1,\ldots,1)}{\partial u_{1}\cdots \partial u_{i-1}}\left /\frac{\partial^{i-1} C(U_{1t},\ldots,U_{i-1,t},1,\ldots,1)}{\partial u_{1}\cdots \partial u_{i-1}} \right.
\label{eq:RosPIT}
\end{equation}
	
For $i=2$, Equation (\ref{eq:RosPIT}) reduces to $V_{1t}=U_{1t}$, $V_{2t}=\partial C(U_{1t},U_{2t})/\partial u_{1}$ because the denominator $ \partial C(U_{1t},1)/\partial u_{1}=1$.
Rosenblatt's transformation has the very convenient property that $\mathbf{U}$ is distributed as copula $\mathbf{C}$ if and only if $\mathcal{R}(\mathbf{U})$ is the \emph{n}-dimensional independent copula
\begin{equation} 
C_{\bot}(\mathbf{V}_t;\hat\theta_t)=\prod_{i=1}^n V_{it} 
\end{equation}
Thus, the Rosenblatt transformation of the original data gives us a vector of \textit{i.i.d.} and mutually independent $Unif(0,1)$ variables, and we can utilize this vector to compare the empirical copula on the transformed data with the independence copula. The KS and CvM tests follow in Equations \ref{eq:kstest1} and \ref{eq:cvmtest1}, respectively. 	
\begin{align}
\mathbf{\hat C}_T (\mathbf{v}) \equiv & \frac{1}{T} \sum_{t=1}^T \prod_{i=1}^n \mathbf{1} \left\{ V_{it}\leq v_{it}\right\} \\ 
KS_R=& \max_t \left | \mathbf{C}_\bot (\mathbf{V}_t;\hat\theta_t)-\mathbf{\hat C}_T (\mathbf{V}_t)\right | \label{eq:kstest1} \\ 
CvM_R=& \sum_{t=1}^T \left\{ \mathbf{C}_\bot (\mathbf{V}_t;\hat\theta_t)-\mathbf{\hat C}_T (\mathbf{V}_t) \right\}^2 \label{eq:cvmtest1}
\end{align}
Critical values of the goodness-of-fit tests are obtained with simulations, as in the \cite{Genest2009} algorithm, as asymptotic distributions are not applicable in the presence of parameter estimation error. In the case of the full-parametric model, the simulations involve generation and estimation of the data from both the model for the margins and for the copula. For the semi-parametric model, the data are generated and estimated only for the copula model. However, as \cite{Patton2012} notes, the approach of combining the non-parametric margins with dynamic copulas does not yet have theoretical support.

Another important issue when working with copulas is the selection of the best copula from the pool. Several methods and tests have been proposed for the selection procedure. The methods proposed by \cite{roncalli00}  are based on the distance from the empirical copula. The authors show how to choose among Archimedian copulas and among a finite subset of copulas. \cite{CHEN2005} propose the use of pseudo-likelihood ratio test for selecting semiparametric multivariate copula models.\footnote{Although some authors use AIC (or BIC) for choosing among two copula models, selection based on these indicators may hold only for the particular sample in consideration (due to their randomness) and not in general. Thus, proper statistical testing procedures are required [see \cite{CHEN2005}].} A test on conditional predictive ability (CPA) is proposed by \cite{Giacomini2006}. This is a robust test that allows one to accommodate both unconditional and conditional objectives. Recently, \cite{Diks:2010fk} have proposed a test for comparing the predictive ability of competing copulas. The test is based on the Kullback-Leibler information criterion (KLIC), and its statistics is a special case of the unconditional version of \cite{Giacomini2006}. 

As our main aim is to employ the model for forecasting, out-of-sample performance of models will be tested by CPA, which consider the forecast performance of two competing models conditional on their estimated parameters to be equal under the null hypothesis
\begin{align}
H_0&: E[ \hat{\mathbf{L}}] = 0 \\
H_{A1}&: E[ \hat{\mathbf{L}}] > 0 \mbox{ and } H_{A2}: E[ \hat{\mathbf{L}}] < 0,
\end{align}
where $\hat{\mathbf{L}} =  \log \mathbf{c}_1(\hat{\mathbf{U}}_,\hat{\gamma}_{1t}) - \log \mathbf{c}_2(\hat{\mathbf{U}}_,\hat{\gamma}_{2t})$. This test can be used for both nested and non-nested models, and we can utilize it for comparison of parametric and semiparametric models as well. The asymptotic distribution of the test statistic is $N(0,1)$, and we compute the asymptotic variance utilizing HAC estimates to correct for possible serial correlation and heteroskedasticity in the differences in log-likelihoods. 

\section{The Data description}

The data set consists of tick prices of crude oil and S\&P 500 futures traded on the platforms of Chicago Mercantile Exchange (CME)\footnote{The data were obtained from the Tick Data, Inc.}. More specifically, oil (Light Crude) is traded on the New York Mercantile Exchange (NYMEX) platform, and the S\&P 500 is traded on the CME in Chicago. We use the most active rolling contracts from the pit (floor traded) session. Prices of all futures are expressed in U.S. dollars.

The sample period spans from January 3, 2003 to December 11, 2012, covering the recent U.S. recession of Dec. 2007 Ð June 2009. We acknowledge the fact that the CME introduced the Globex$\textregistered$ electronic trading platform on Monday, December 18, 2006, and begun to offer nearly continuous trading. However, we restrict the analysis on the intraday 5-minutes returns within the business hours of the New York Stock Exchange (NYSE) as most of their liquidity of S\&P 500 futures comes from the period when U.S. markets are open. Time synchronization of our data is achieved in a way that oil prices are paired with the S\&P 500 by matching the identical Greenwich Mean Time (GMT) stamps. We eliminate transactions executed on Saturdays and Sundays, U.S. federal holidays, December 24 to 26, and December 31 to January 2 because of the low activity on these days, which could lead to estimation bias. Hence, in our analysis we work with data from 2,436 trading days. 
\begin{figure}
\begin{center}
\includegraphics[width=0.47\textwidth]{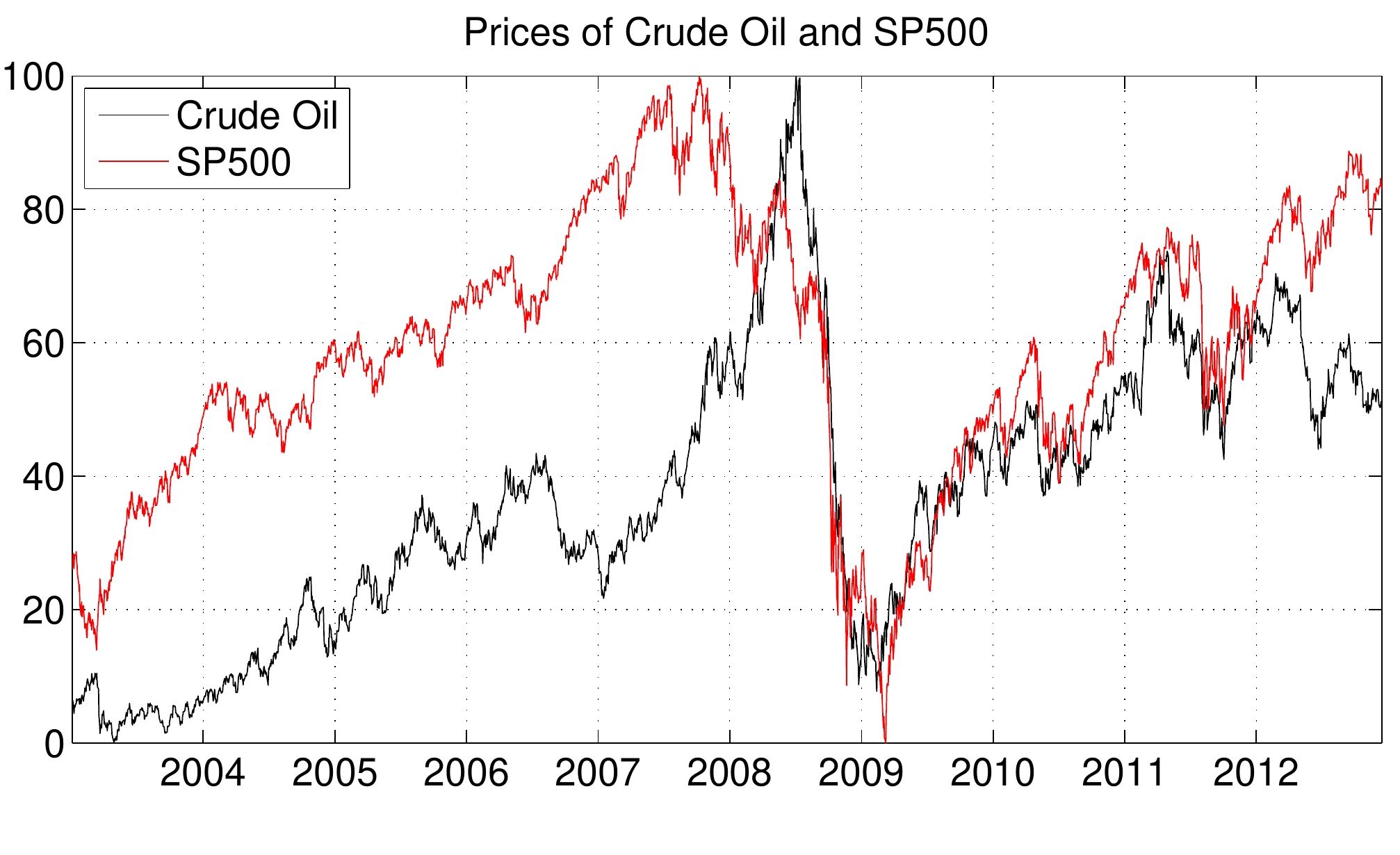}
\includegraphics[width=0.47\textwidth]{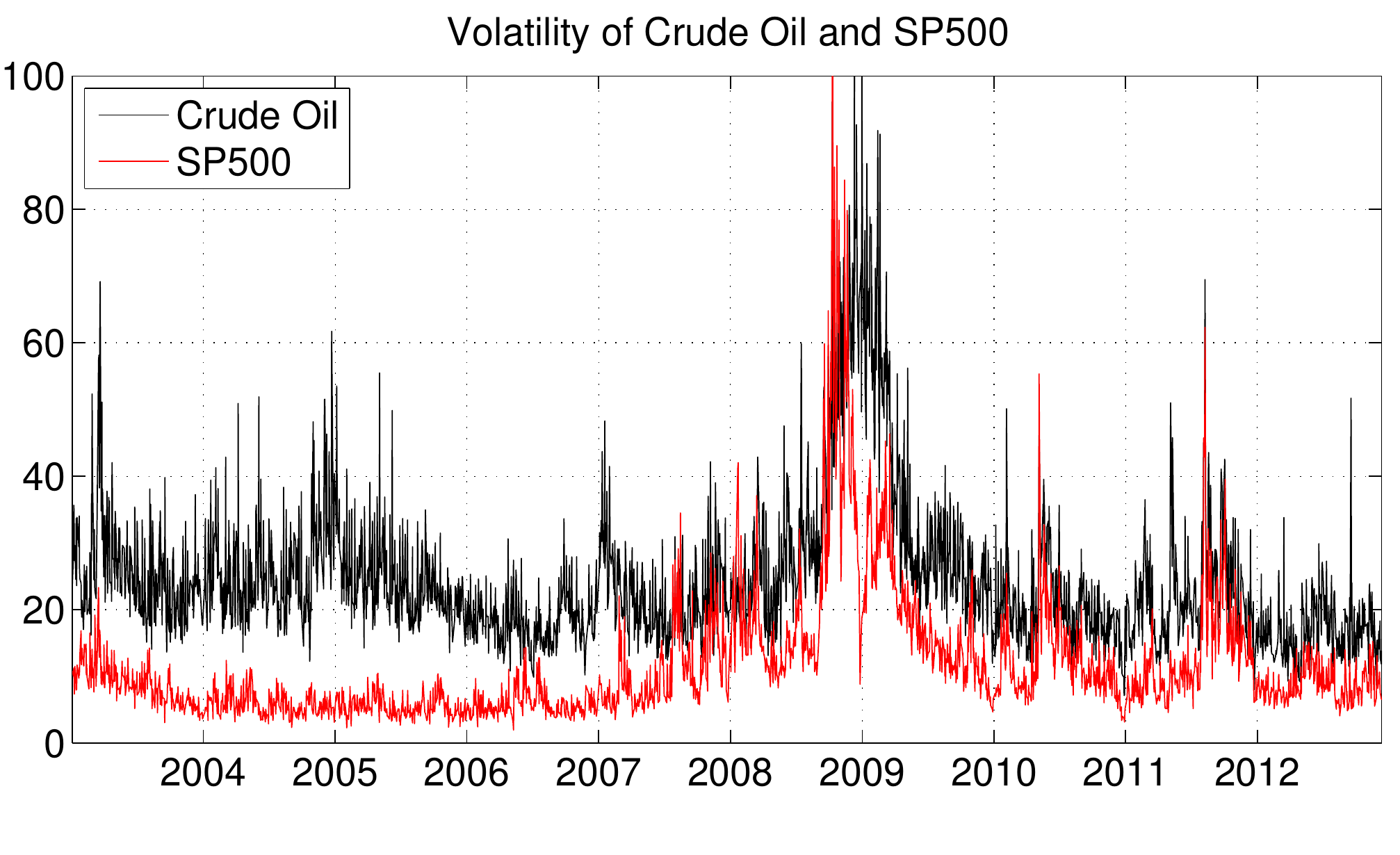}
\end{center}
\caption{Normalized prices and annualized realized volatilities of oil and stocks (S\&P 500), over the sample period extending from January 3, 2003 to December 11, 2012.}
\label{fig:prices}
\end{figure}

For our empirical model, we need two time series, namely daily returns and realized variance to be able to estimate the realized GARCH model in margins. We consider open-close returns; thus, daily returns are simply obtained as a sum of logarithmic intraday returns. Realized variance is computed as a sum of squared 5-minute intraday returns
\begin{equation}
RV_t = \sum_{i=1}^M r_i^2,
\end{equation}
Figure \ref{fig:prices} plots the development of prices of the oil and stocks together with its realized volatility. Please note that plot of prices is normalized to make them comparable and for the plot of realized volatility, we utilize daily volatility annualized according to the following convention: $100\times\sqrt{250\times RV_t}$. Strong time-varying nature of the volatility can be noticed immediately for both oil and stocks. In Table \ref{tab:sumStat}, we present descriptive statistics of the returns of the data that constitute our sample. Distributions of the daily returns are showing excess kurtosis. It is interesting that the volatility of oil is on average more than twice larger in comparison to the volatility of stocks. In addition, the dynamics of volatilities differ, primarily in the first part of the period. This, in fact, motivates a need for the flexible model, which will capture the different dynamics in the marginal distributions for oil and stocks. 

\section{Empirical Results}
Before modeling the dependence structures between oil and stocks, we must model their conditional marginal distributions first. Utilizing the Bayesian Information Criterion (BIC) and considering general ARMA models up to five AR as well as MA lags, we find an AR(2) model to best capture time-varying dependence in the mean of S\&P 500 stock market returns, while no significant dependence in the mean was found for oil.

Table \ref{tab:condVarModel} summarizes the in-sample realized GARCH(1,1) fit for both oil and stocks represented by the S\&P 500 in our study. In addition, the benchmark volatility model from the GARCH family, namely, GJR model \citep{gjr93}, is used for comparison. All the estimated parameters are significantly different from zero and are similar to those obtained by \cite{Hansen2012}. We can see that realized volatility plays its role in the model as it helps to model volatility significantly. By observing partial log-likelihood $\mathcal{LL}_{r}$ as well as information criteria, we can see that the realized GARCH brings significant improvement over the GJR GARCH model in both oil and stocks. This is a crucial result for copulas, as we need to specify the best possible model in the margins to make sure there is no univariate dependence left. If a misspecified model is utilized for the marginal distributions, then the probability integral transforms will not be $Unif(0,1)$ distributed, and this will result in copula misspecification. 
\begin{table}[h] 
\footnotesize
\begin{tabular}{lrrrrlrrrr}\toprule 
 & \multicolumn{2}{c}{Crude Oil}& \multicolumn{2}{c}{S\&P 500}&&   \multicolumn{2}{c}{Crude Oil}& \multicolumn{2}{c}{S\&P 500}\\ 
\midrule 
& \multicolumn{4}{c}{AR(2)}&&\multicolumn{4}{c}{AR(2)}\\
\cmidrule{2-5} \cmidrule{7-10}  
$c$ &   0.0001 &(0.29)& 0.0000 &(0.21)&$c$  & 0.0001 &(0.29)& 0.0000 &(0.21)\\ 
$\alpha_1$  &- &- & -0.1095 &(-5.42)&$\alpha_1$ & - &- & -0.1095 &(-5.42)\\ 
$\alpha_2$   &- &- & -0.0744 &(-3.68)&$\alpha_2$   &- &- & -0.0744 &(-3.68)\\ 
\cmidrule{2-5} \cmidrule{7-10} 
& \multicolumn{4}{c}{realized GARCH(1,1)}&&\multicolumn{4}{c}{GJR GARCH(1,1)}\\
\cmidrule{2-5} \cmidrule{7-10}  
$\omega$  & 0.0626 &(6.14)& 0.2000 &(14.07)&$\kappa$  & 0.0143 &(2.69)& 0.0028 &(2.79)\\ 
$\beta$  & 0.7622 &(46.41)& 0.7176 &(45.11)&$\phi$  & 0.0270 &(2.57)& 0.0187 &(1.71)\\ 
$\gamma$  & 0.2081 &(12.59)& 0.2413 &(19.04)&$\iota$  & 0.0390 &(2.61)& 0.0883 &(5.72)\\ 
$\xi$  & -0.3173 &(-9.26)& -0.9018 &(-21.84)&$\psi$  & 0.9363 &(72.58)& 0.9321 &(85.86)\\ 
$\phi$  & 1.0758 &(23.36)& 1.1130 &(40.87)& &-&-&-&-\\ 
$\tau_1$  & -0.0627 &(-7.18)& -0.0772 &(-8.15)& &-&-&-&-\\  
$\tau_2$  & 0.1053 &(16.62)& 0.0999 &(16.56)& &-&-&-&-\\ 
$\nu$  & 13.4633 &(4.07)& 12.2552 &(5.49)&$\nu$  & 12.7026 &(4.29)& 7.9716 &(6.48)\\ 
$\lambda$  & -0.0885 &(-3.21)& -0.1544 &(-6.37)& &-&-&-&-\\ 
\cmidrule{2-5} \cmidrule{7-10}  
$\mathcal{LL}_{r,x}$  &\multicolumn{2}{c}{    -4558.16 } &\multicolumn{2}{c}{    -4167.49 } & &\multicolumn{2}{c}{    - }& \multicolumn{2}{c}{    - }\\ 
$\mathcal{LL}_{r}$  &\multicolumn{2}{c}{  -3189.22 } &\multicolumn{2}{c}{    -2473.56 } &$\mathcal{LL}$  &\multicolumn{2}{c}{    -3207.74 } &\multicolumn{2}{c}{    -2501.89 } \\ 
$AIC_r$  &\multicolumn{2}{c}{     6396.43 } &\multicolumn{2}{c}{     4965.11 }&$AIC$   &\multicolumn{2}{c}{     6425.49 } &\multicolumn{2}{c}{     5013.78 } \\ 
 
$BIC_r$   &\multicolumn{2}{c}{    6448.61 } &\multicolumn{2}{c}{   5017.30 } &$BIC$   &\multicolumn{2}{c}{     6454.48 } &\multicolumn{2}{c}{     5042.77 } \\ 
  \bottomrule  
\end{tabular}   
\caption{Parameter estimates from AR($2$) \emph{log-linear} realized GARCH(1,1) and benchmark GJR GARCH(1,1), the former with \emph{skew-t} innovations and the latter with standard StudentÕs \emph{t}. \emph{t}-statistics are reported in parentheses.}  
\label{tab:condVarModel}  
\end{table} 

For the estimated standardized residuals from the realized GARCH(1,1), we consider both parametric and nonparametric distributions, as noted previously. The figure in \ref{fig:fittedMargins} plots the histogram of the standardized residuals together with quantile plots against skewed $t$ distribution. We can see a reasonable fit of skewed $t$ distribution with the data, although a very small departure from tails can be observed for the S\&P 500 data. This also motivates us to choose to estimate a full battery of copula models including those combining a nonparametric empirical distribution for margins and a parametric copula function; although from the density fits, we can see that the gains will probably not be large. 

 To study the goodness of fit for the skewed $t$ distribution, we compute the Kolmogorov-Smirnov (KS) and Cramer-von Mises (CvM) test statistics with $p$-values from 1,000 simulations, and we find KS (CvM) $p$-values of 0.452 (0.577) and 0.254 (0.356) for the oil and S\&P 500 standardized residuals, respectively. Thus, we are not able to reject the null hypothesis that these distributions come from the skewed $t$, which provides support for these models of the marginal distribution. The estimated parameters $\nu$ ($\lambda$) for the oil and stocks are 13.462 (-0.088) and 12.255 (-0.154), respectively. This allows us to continue with modeling time-varying dependence. 
 
 \subsection{Time-varying dependence between oil and stocks}
 
 By studying simple correlation measures of original returns, we find the linear correlation and rank correlation for oil and stocks to be 0.29 and 0.224, respectively, both significantly different from zero. Before specifying a functional form for a time-varying copula function, we test for the presence of time-varying dependence utilizing the simple approach based on the ARCH LM test. The test statistics are computed from the OLS estimate of the covariance matrix, and critical values are obtained employing \textit{i.i.d.} bootstrap (for detailed information, consult \cite{Patton2012}). Computing the test for the time-varying dependence between oil and stocks up to $p=10$ lags, we find the joint significance of all coefficients. Thus, we can conclude that there is evidence against constant conditional correlation for oil and stocks.

\begin{table}
\footnotesize
\begin{center} 
\begin{tabular*}{\textwidth}{@{\extracolsep{\fill}}llcccccccc}
\toprule 
& &  \multicolumn{3}{c}{Parametric} & & \multicolumn{3}{c}{Semiparametric}\\ 
\cmidrule{2-5} \cmidrule{7-10}  
		& & & & \\
 \multicolumn{4}{l}{\bf{Constant copula}} \\ 
 	& & \multicolumn{2}{c}{Est. Param} & $\mathcal{\log L}$ & & \multicolumn{2}{c}{Est. Param} & $\mathcal{\log L}$ & \\
	\cmidrule{3-5} \cmidrule{7-10} 
Normal & 		$\rho$ &  0.2060 & (0.0290) & \textbf{52.56 } & &   0.2053 & (0.0231) & \textbf{       52.43 } \\ 
		\cmidrule{3-5} \cmidrule{7-10} 
Clayton  &$\kappa$ &  0.2392 &(0.0353) & \textbf{       56.90 }&  &0.2738 &(0.0322) & \textbf{       58.69 }\\ 
		\cmidrule{3-5} \cmidrule{7-10} 
RGumb & $\kappa$ &  1.1403 &(0.0213) & \textbf{       66.40 }  & &  1.1588 &(0.0176) &  \textbf{       69.13 }\\ 
		\cmidrule{3-5} \cmidrule{7-10} 
Student's $t$ & $\rho$ &  0.2051 &(0.0261) &  &&0.2183 &(0.0214) & \\ 
		   & $\nu^{-1}$ &  0.1376 &(0.0244) &   \textbf{       79.74 } &&0.1660 &(0.0252) & \textbf{       81.78 } \\ 
		\cmidrule{3-5} \cmidrule{7-10} 
Sym. Joe-Clayton		& $\tau^L$ &  0.0941 &(0.0268)&  && 0.1209 &(0.0254)\\ 
		& $\tau^U$ &  0.0208 &(0.0226)&  \textbf{       66.14 } && 0.0242&(0.0193) &  \textbf{       67.38 } \\ 	
		& & & & \\
 \multicolumn{4}{l}{\bf{``GAS" time-varying copula}} \\ 
  	& & \multicolumn{2}{c}{Est. Param} & $\mathcal{\log L}$ & & \multicolumn{2}{c}{Est. Param} & $\mathcal{\log L}$ & \\
	\cmidrule{3-5} \cmidrule{7-10} 
$RGumb_{GAS}$ 	& $\hat{\omega}$ &  -0.0074 &  (0.2071) & & &  -0.0097 & (0.4219) & \\ 
				&$\hat{\alpha}$ &  0.1038 & (0.3131) &  & & 0.1184 & (0.3575) & \\ 
 				&$\hat{\beta}$ &  0.9972 & (0.0122) & \textbf{135.06 } & &  0.9960 & (0.0447) & \textbf{139.15 }\\ 
\cmidrule{3-5} \cmidrule{7-10} 
$N_{GAS}$	&$\hat{\omega}$ &  0.0017 & (0.0037) && &  0.0019 & (0.0041) &  \\ 
				&$\hat{\alpha}$ &  0.0474 & (0.0109) & & &  0.0553 & (0.0124) & \\ 
				&$\hat{\beta}$ &  0.9952 & (0.0070) &  \textbf{      152.47 } & &  0.9947 & (0.0075) &  \textbf{      153.59 }\\ 
\cmidrule{3-5} \cmidrule{7-10} 
$t_{GAS}$	&$\hat{\omega}$ &  0.0016 & (0.0040) & & &  0.0018 & (0.0073) &  \\ 
				&$\hat{\alpha}$ &  0.0493 & (0.0128) & & &  0.0579 & (0.0193) & \\ 
				&$\hat{\beta}$ &  0.9957 & (0.0076) & & &  0.9952 & (0.0205) &  \\ 
				&$\hat\nu^{-1}$ &  0.0775 & (0.0259) & \textbf{      162.82 }& &  0.0940 & (0.0315) &  \textbf{      165.39 } \\ 
  \bottomrule  
\end{tabular*}   
\end{center} 
\caption{ Constant and time-varying copula model parameter estimates with AR($2$) realized GARCH(1,1) model for both fully parametric and semiparametric cases. Bootstrapped standard errors are reported in parentheses.}  
\label{tab:UnivRGTvOil-SP500}  
\end{table} 

Motivated by the possible time-varying dependence in oil and stocks, we can specify the copula functions. We estimate three time-varying copula functions, namely, Normal, rotated Gumbel and Student's $t$ using the GAS framework described in the methodology part. As a benchmark, we also estimate the constant copulas to be able to compare the time-varying models against the constant ones. While semiparametric approach is empirically interesting and not often used in literature, we employ it for all the estimated models as well.

Table \ref{tab:UnivRGTvOil-SP500} presents the fit from all estimated models. Starting with constant copulas, all the parameters are significantly different from zero, and StudentÕs $t$ copula appears to describe the oil and stock pair best according to highest log-likelihood. Semiparametric specifications combining nonparametric distribution in margins with parametric copula function bring further improvement in the log-likelihoods. Importantly, time-varying specifications bring large improvement in log-likelihoods and confirm strong time-varying dependence between oil and stocks. 

To study the goodness of fit for all the specified models, we utilize\footnote{The results of the in-sample goodness of fit tests are available on request from the authors. We do not include them in the text to save space.} Kolmogorov-Smirnov (KS) and Cramer-von Mises (CvM) test statistics with $p$-values obtained from 1,000 simulations. The methodology is described in detail in previous sections.  None of the fully parametric models is rejected, while most of the semiparametric models are rejected with exception of constant StudentÕs $t$, Sym. Joe-Clayton and time-varying StudentÕs $t$. This result suggests that fully parametric models with realized GARCH and parametric distribution in margins are all well specified. Thus, the realized GARCH appears to model very well all the dependence in margins, which is crucial for the good specification of the model in the copula-based approach. Semiparametric models are interestingly rejected and are not specified well, except for a few mentioned cases. This is in line with results of \cite{Patton2012}, who finds rejections in semiparametric specifications in the U.S. stock indices data. Still, both tests strongly support the realized GARCH time-varying GAS copulas for the oil and stock pair.

\subsection{Out-of-sample comparison of the proposed models}

While it is important to have a well-specified model that describes the data, most of the times we are interested in utilizing the model in predictions. Thus, we conduct an out-of-sample evaluation of the proposed models. For this, the sample is divided into two periods. The first, the in-sample period, is used to obtain parameter estimates from all models and spans from January 3, 2003 to July 6, 2010. The second, the out-of-sample period, is then used for evaluation of forecasts. Due to highly computationally intensive estimations of the models, we restrict ourselves to a fixed window evaluation, where the models are estimated only once, and all the forecasts are performed using the recovered parameters from this fixed in-sample period. This makes it even harder for the models to perform well in the highly dynamic data.

\begin{table} 
\footnotesize
\begin{tabular*}{\textwidth}{@{\extracolsep{\fill}}lllllllll}
\toprule 
& \multicolumn{8}{c}{Parametric margins}\\ 
\cmidrule{2-9} 
                      &       Normal  &      Clayton  &      R. Gum.  &       Stud. $t$  &          SJC  & $RGum_{GAS}$  &    $N_{GAS}$  &    $t_{GAS}$  \\ \cmidrule{2-         9}
Normal                &            &            &            &            &            &            &            &            \\
Clayton               &         0.36  &            &            &            &            &            &            &            \\
R. Gum.               &         $2.00^{\ast\ast}$  &         $4.37^{\ast\ast\ast}$  &            &            &            &            &            &            \\
Stud. $t$                &         $1.78^{\ast}$  &         $2.09^{\ast\ast}$  &        -0.15  &            &            &            &            &            \\
SJC                   &         $1.91^{\ast}$  &         $3.80^{\ast\ast\ast}$  &        -1.94  &        -0.68  &            &            &            &            \\
$RGum_{GAS}$          &        $ 2.75^{\ast\ast\ast}$  &         $3.09 ^{\ast\ast\ast}$ &         $2.34^{\ast\ast\ast}$  &         $1.99^{\ast\ast}$  &         $2.53^{\ast\ast\ast}$  &            &            &            \\
$N_{GAS}$             &         $2.49^{\ast\ast\ast}$  &        $ 2.31^{\ast\ast}$  &         $1.65^{\ast}$  &         1.48  &         $1.94^{\ast}$  &         0.33  &            &            \\
$t_{GAS}$             &         $3.94^{\ast\ast\ast}$  &         $4.02^{\ast\ast\ast}$  &         $3.42^{\ast\ast\ast}$  &         $3.27^{\ast\ast\ast}$  &         $3.73^{\ast\ast\ast}$  &         $2.37^{\ast\ast\ast}$  &         1.09  &            \\
\cmidrule{2-9} 
$\mathcal{LL}^{OOS}$  &        33.14  &        34.62  &        43.74  &        43.17  &        40.74  &        60.10  &        62.65  &        69.79  \\
Rank                  &         8.00  &         7.00  &         4.00  &         5.00  &         6.00  &         3.00  &         2.00  &         1.00  \\
\cmidrule{2-9} 
& \multicolumn{8}{c}{Nonparametric margins}\\ 
\cmidrule{2-9} 
                      &       Normal  &      Clayton  &      R. Gum.  &       Stud. $t$  &          SJC  & $RGum_{GAS}$  &    $N_{GAS}$  &    $t_{GAS}$  \\ \cmidrule{2-         9}
Normal                &            &            &            &            &            &            &            &            \\
Clayton               &         1.12  &            &            &            &            &            &            &            \\
R. Gum.               &         $2.23^{\ast\ast}$  &         $3.88^{\ast\ast\ast}$  &            &            &            &            &            &            \\
Stud. $t$                &         $2.17^{\ast\ast}$  &         $1.93^{\ast}$  &         0.43  &            &            &            &            &            \\
SJC                   &         $2.39^{\ast\ast\ast}$  &         $3.77^{\ast\ast\ast}$  &        -1.18  &        -0.74  &            &            &            &            \\
$RGum_{GAS}$          &        $ 3.24^{\ast\ast\ast}$  &         $3.34^{\ast\ast\ast}$  &         $2.90^{\ast\ast\ast}$  &        $ 2.35^{\ast\ast\ast}$  &        $ 2.95^{\ast\ast\ast}$  &            &            &            \\
$N_{GAS}$             &         $2.64^{\ast\ast\ast}$  &         $2.32^{\ast\ast\ast}$  &         $1.90^{\ast}$  &         1.57  &         $2.03^{\ast\ast}$  &         0.13  &            &            \\
$t_{GAS}$             &         $4.05^{\ast\ast\ast}$  &         $3.92^{\ast\ast\ast}$  &        $ 3.55^{\ast\ast\ast}$  &         $3.20^{\ast\ast\ast}$  &         $3.69^{\ast\ast\ast}$  &         $1.97^{\ast\ast}$  &         1.06  &            \\
\cmidrule{2-9} 
$\mathcal{LL}^{OOS}$  &        28.74  &        32.73  &        38.51  &        39.96  &        37.42  &        59.11  &        60.04  &        66.49  \\
Rank                  &         8.00  &         7.00  &         5.00  &         4.00  &         6.00  &         3.00  &         2.00  &         1.00  \\
\cmidrule{2-9} 
& \multicolumn{8}{c}{Parametric vs. nonparametric margins}\\ 
\cmidrule{2-9} 
               &       Normal  &      Clayton  &      R. Gum.  &       Stud t  &          SJC  & $RGum_{GAS}$  &    $N_{GAS}$  &    $t_{GAS}$  \\ \cmidrule{2-         9}
\emph{t-stat}  &         0.85  &         0.77  &         0.90  &         0.83  &         0.82  &         0.73  &         0.78  &         0.83  \\

  \bottomrule  
\end{tabular*}   
{\tiny{$\ast$, $\ast\ast$ and $\ast\ast\ast$ denote significantly better performance at the 90\%, 95\% and 99\% significance levels, respectively.}}\\
\caption{The t-statistics from the out-of-sample pair-wise comparisons of log-likelihood values for five constant copula models and three time-varying copula models, with fully parametric or semiparametric marginal distribution models. Positive (negative) values indicate better performance of copula in the row (column) to a copula in the column (row). $\mathcal{LL}^{OOS}$ is the out-of-sample log likelihood, and ``Rank" simply ranks all the models with respect to the log likelihood. In the bottom row, we compare the performance of the same copula with different margins \emph{i.e.} parametric vs. nonparametric ones. The out-of-sample period is from July 6, 2010 to December 11, 2012 and includes 609 observations. } 
\label{tab:UnivRGOosCompOil-SP500} 
\end{table} 
 
For the out-of-sample forecast evaluation, we employ the conditional predictive ability (CPA) test of \cite{Giacomini2006}. Table \ref{tab:UnivRGOosCompOil-SP500} presents the results from this test. The time-varying copula models significantly outperform the constant copula models in the out-of-sample evaluation. This holds both for the parametric and semiparametric cases. Thus, time-varying copulas have much stronger support for forecasting the dynamic distribution of oil and stocks. When comparing the different time-varying copula functions, the test is not so conclusive. While StudentÕs $t$ statistically outperform rotated Gumbel, the forecasts from StudentÕs $t$ can not be statistically distinguished from the normal copula. When looking at the out-of-sample log-likelihoods, Student's $t$ copula is the most preferred. Finally, the bottom row shows that forecasts from parametric models and semiparametric ones cannot be statistically distinguished.
 
Thus, we find strong statistical support that the realized GARCH time-varying copula methodology well describes the dynamic joint distribution of the oil and stocks in both the in-sample and out-of-sample. 

\subsection{Time-varying correlations and tails}
\begin{figure}
\begin{center}
\includegraphics[width=0.45\textwidth]{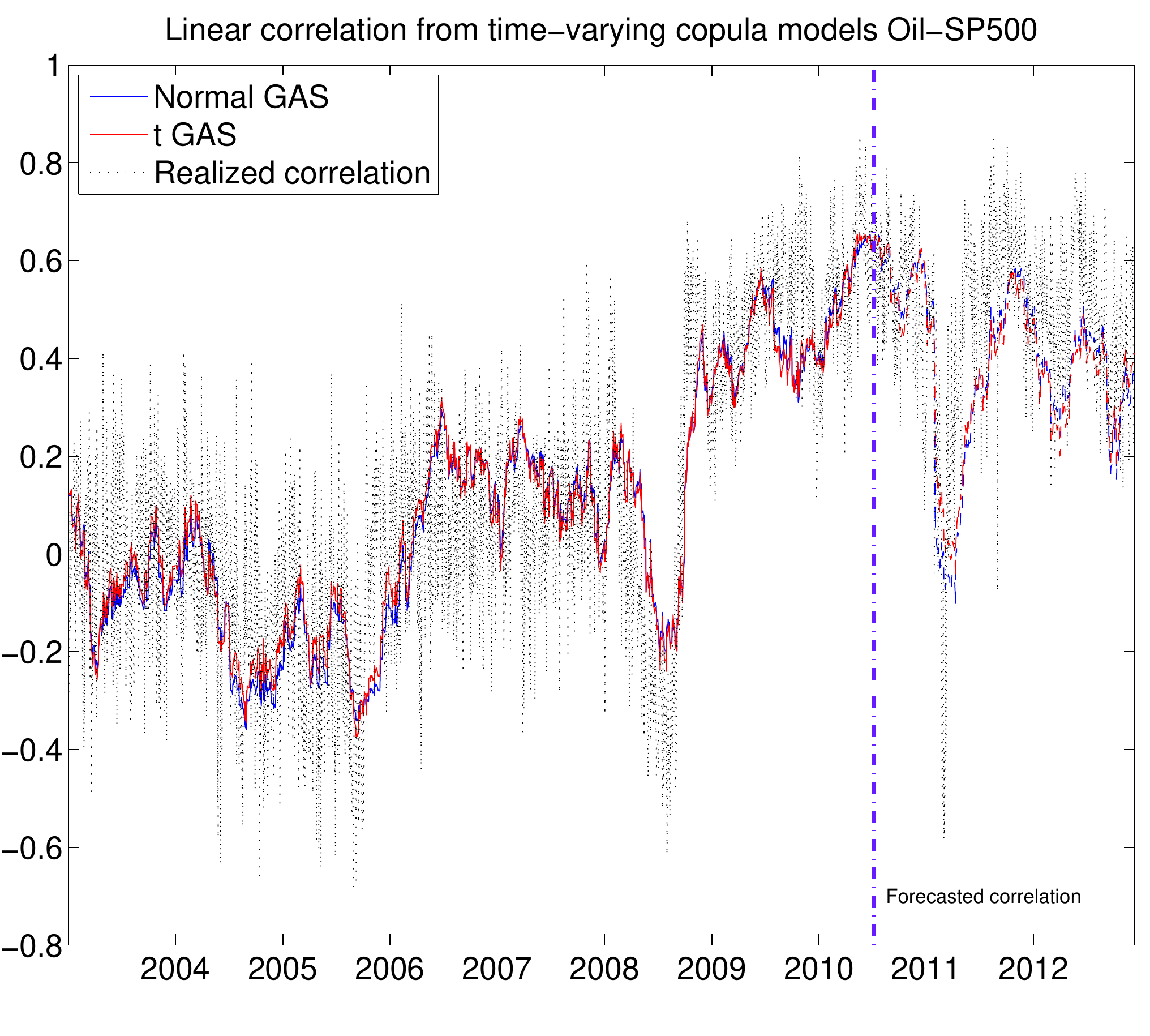}
\includegraphics[width=0.46\textwidth]{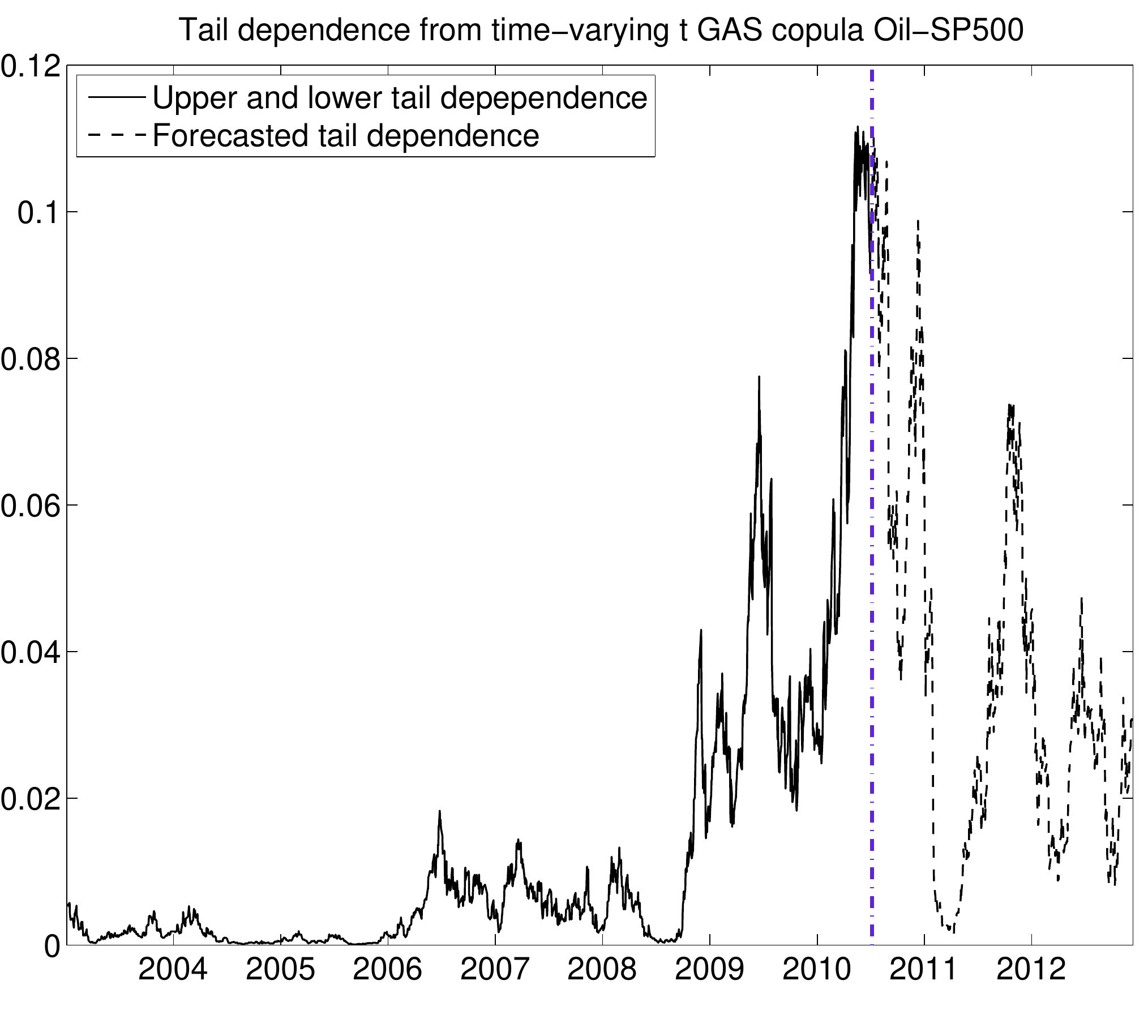}
\end{center}
\caption{Left: linear correlation plotted against realized correlation. Right: tail dependence from time-varying Student's \emph{t}-copula. The vertical dashed line separates the in-sample from the out-of-sample (forecasted) part. }
\label{fig:correlations}
\end{figure}

Having correctly specified the empirical model capturing the dynamic joint distribution between oil and stocks, we can proceed to studying the pair. Figure \ref{fig:correlations} plots the time-varying correlations implied by our model with normal and StudentÕs $t$ GAS copulas. In the first period, correlations are obtained from an in-sample fit of our model. In the second period, the model is used to forecast the correlations. These dynamics are very close to the one reported by a recent study of \cite{wen2012measuring}, although they are more accurate due to the help of realized measures utilized in our modeling strategy and GAS structure as well. To highlight this point, we compare the in-sample and forecasted correlations from the model to the actual correlation measured with the help of high-frequency data. The realized correlation between oil and stocks for a given day $t$ is measured using $k=1,\ldots,M$ five-minute returns non-parametrically using realized volatility \citep{abdl2003,barndorff2004b} as 
$$RCorr_{t}=\frac{\sum_{k=1}^M r_{(1)k,t}r_{(2)k,t}}{\sqrt{\sum_{k=1}^M r_{(1)k,t}^2}\sqrt{\sum_{i=1}^M r_{(2)k,t}^2}},$$
and is depicted by the black dotted line in the Figure \ref{fig:correlations}. We note that realized correlations provide noisy estimates, and we can clearly see how correlations implied by our model fit the actual correlations, as they also capture abrupt change in the year 2008. The out-of-sample forecast of the correlations lags the actual correlations and is slightly downward biased. The reason for this departure is that we use a static forecast utilizing parameters estimated on the in-sample data set as explained earlier in the text. Rolling sample forecasts would recover the actual correlations with much better precision. Hence, the proposed approach can capture the time-varying dynamics very well and is also able to forecast the dependence. 

As we can see from the Figure \ref{fig:correlations}, the dependence between oil and stocks varies strongly over time. \cite{wen2012measuring} suggests that the correlations changed dramatically during the 2008 crisis, but employing a larger data span, we suspect the correlation to have more regimes. To find the presence of structural breaks statistically, we employ the $supF$ test 
\citep{hansen1992tests,andrews1993tests}, with p-values computed based on \cite{hansen1997approximate} and apply it to the correlations.\footnote{To conserve space, we do not report the test statistics for the detection of structural breaks. The results are available upon request. The test is used to evaluate the null hypothesis of no structural change, utilizing an extension of the $F$ test statistics. In the first step, the error sum of squares is computed together with the restricted sum of the squares for every potential change point utilizing least squares fits. Second, $F$ test statistics are computed for every potential change point, and third, a supremum is found from all the $F$ test statistics constituting the structural break. The p-values are computed based on \cite{hansen1997approximate}.} Employing this approach, we confirm two endogenous changes in the dependence, with March 14, 2006 and October 9, 2008 dividing the data into three distinct periods. 

During the first period, the correlation was decreasing from zero to negative values. An economic reasoning for this finding comes from the fact that the response of the stock markets to oil price shocks differs according to the origin of such shocks \citep{hamilton1996happened}. Specifically, a supply-side shock negatively impacts stock market returns and leads to negative correlation in the oil-stocks pair. An increase in oil prices -- a supply-side shock -- might result from an abrupt reduction of output by major producers (e.g., OPEC countries) or due to a major political event, such as the 1990-1991 Gulf War. 

During the second period, beginning March 14, 2006, the correlation increased to positive values, while after the turmoil of October 2008, identified by the test very precisely with the date of October 9, 2008, the correlation became significantly positive, suggesting that diversification opportunities are disappearing. In the following years, correlations remained high, while in the last years of the sample, they began to decrease but remained in positive territory. This may be attributed to the changing expectations of market participants. After the financial crisis, the oil market became very strongly financialized \citep{buyukcsahin2014speculators}, and moves in stock prices also appeared to carry over to oil prices as well. After 2008, stock market participants were much more uncertain about future behavior, which translated into high volatility during this period.

In addition, the second part of the Figure \ref{fig:correlations} plots the dynamic tail dependence from the StudentÕs $t$ GAS model. In the first period, we document tail independence. The 2008 turmoil brought a large increase in tail dependence, which remained highly dynamic. While one of the additional advantages time-varying copula functions bring is allowing for asymmetric tails; in the final empirical model, we also study tail asymmetry. Interestingly, we find no evidence for asymmetry in the tails.

Our results have serious implications for investors as they suggest that diversification possibilities may be even larger than commonly perceived from the mere dynamics of the correlations. In addition, correlations as well as tail dependence have been rapidly changing over the past few years. We will utilize the results and study the possible economic benefits of our analysis, with the main focus on translating the dynamic correlation and tail dependence to a proper quantification of the diversification benefits. 

\section{Economic implications: Time-varying diversification benefits and VaR}

Statistically significant improvement in the fit, or even out-of-sample forecasts does not necessarily need to translate into economic benefits. Thus, we test the proposed methodology in economic implications. First, we quantify the risk of an equally weighted portfolio composed from oil and stocks, and second, we study the benefits from diversification to see how the strongly varying correlation affects the diversification benefits.

\subsection{Quantile forecasts}
\begin{table}
\tiny
\caption{Out-of-sample VaR evaluation for GJR GARCH and realized GARCH models in margins. Empirical quantile $\hat{C}_{\alpha}$, estimated Giacomini and Komunjer (2005) $\hat{L}$, logit DQ statistics and its 1000$\times$ simulated $p$-val are reported. $\hat{L}$ is moreover tested with Diebold-Mariano statistics with the Newey-West estimator for variance. All models are compared to $t_{GAS}$ with realized GARCH in margins, while models with significantly less accurate forecasts at 95\% level are reported in bold.}
\centering
\begin{tabular*}{1.1\textwidth}{@{\extracolsep{\fill}}llrrrrrrrrrrrrrr}
\toprule 
 & &\multicolumn{6}{c}{Parametric} & & \multicolumn{6}{c}{Semiparametric}  \\ 
\cmidrule{3-8}\cmidrule{10-15}
& & 0.01 & 0.05 & 0.1 & 0.9 & 0.95 & 0.99 & &  0.01 & 0.05 & 0.1 & 0.9 & 0.95 & 0.99\\  
\cmidrule{1-2} \cmidrule{3-8}\cmidrule{10-15} 
\multirow{24}{*}{\rotatebox[origin=c]{90}{GJR GARCH}} & \textbf{Normal} \\
&$\hat{C}_{\alpha}$&0.015 & 0.069 & 0.110 & 0.893 & 0.946 & 0.995 & & 0.015 & 0.071 & 0.110 & 0.893 & 0.946 & 0.993 \\
 &$\hat{L}$ &0.025 & \textbf{0.083} & 0.132 & \textbf{0.108} & \textbf{0.062} & 0.015 & & 0.024 & 0.082 & 0.131 & \textbf{0.108} & \textbf{0.061} & 0.015 \\
 &DQ &8.392 & 8.031 & 4.323 & 3.740 & 8.755 & 3.572 & & 8.392 & 8.468 & 3.126 & 7.352 & 4.962 & 1.674 \\
 &$p$-val &0.211 & 0.236 & 0.633 & 0.712 & 0.188 & 0.734 & & 0.211 & 0.206 & 0.793 & 0.290 & 0.549 & 0.947 \\
 &\\
&$\mathbf{RGumb_{GAS}}$ \\
 &$\hat{C}_{\alpha}$&0.008 & 0.043 & 0.090 & 0.901 & 0.957 & 0.998 & & 0.008 & 0.039 & 0.087 & 0.905 & 0.959 & 0.998 \\
 &$\hat{L}$ &0.024 & 0.082 & 0.132 & \textbf{0.107} & \textbf{0.061} & 0.016 & & 0.024 & 0.082 & 0.132 & \textbf{0.107} & \textbf{0.061} & 0.016 \\
 &DQ &5.779 & 6.305 & 3.911 & 10.334 & 5.496 & 6.494 & & 5.779 & 6.831 & 3.260 & 9.267 & 6.232 & 6.494 \\
 &$p$-val &0.448 & 0.390 & 0.689 & 0.111 & 0.482 & 0.370 & & 0.448 & 0.337 & 0.776 & 0.159 & 0.398 & 0.370 \\
 &\\
&$\mathbf{N_{GAS}}$ \\
 &$\hat{C}_{\alpha}$&0.013 & 0.043 & 0.090 & 0.911 & 0.967 & 0.998 & & 0.010 & 0.041 & 0.087 & 0.913 & 0.970 & 0.998 \\
 &$\hat{L}$ &0.024 & 0.083 & 0.133 & 0.106 & \textbf{0.061} & \textbf{0.016} & & 0.024 & \textbf{0.083} & \textbf{0.133} & 0.107 & \textbf{0.062} & \textbf{0.017} \\
 &DQ &4.486 & 4.284 & 3.869 & 7.479 & 9.155 & 6.494 & & 4.838 & 4.667 & 5.314 & 10.466 & 13.575 & 6.494 \\
 &$p$-val &0.611 & 0.638 & 0.694 & 0.279 & 0.165 & 0.370 & & 0.565 & 0.587 & 0.504 & 0.106 & 0.035 & 0.370 \\
 &\\
&$\mathbf{t_{GAS}}$ \\
&$\hat{C}_{\alpha}$&0.010 & 0.046 & 0.085 & 0.908 & 0.966 & 0.998 & & 0.011 & 0.044 & 0.089 & 0.905 & 0.969 & 0.998 \\
 &$\hat{L}$ &0.024 & 0.082 & 0.133 & 0.107 & \textbf{0.061} & \textbf{0.017} & & 0.024 & 0.082 & 0.133 & 0.107 & \textbf{0.062} & \textbf{0.017} \\
 &DQ &4.838 & 3.072 & 3.610 & 8.370 & 9.413 & 6.494 & & 9.276 & 5.442 & 3.167 & 11.390 & 11.982 & 6.494 \\
 &$p$-val &0.565 & 0.800 & 0.729 & 0.212 & 0.152 & 0.370 & & 0.159 & 0.488 & 0.788 & 0.077 & 0.062 & 0.370 \\
\cmidrule{1-2} \cmidrule{3-8}\cmidrule{10-15} 
&\\
\multirow{24}{*}{\rotatebox[origin=c]{90}{realized GARCH}} &\textbf{Normal} \\
 &$\hat{C}_{\alpha}$ &0.023 & 0.082 & 0.130 & 0.877 & 0.931 & 0.987 & & 0.021 & 0.085 & 0.126 & 0.878 & 0.931 & 0.985 \\
 &$\hat{L}$ &\textbf{0.026} & \textbf{0.083} & 0.132 & \textbf{0.107} & \textbf{0.062} & 0.015 & & 0.025 & \textbf{0.083} & 0.132 & \textbf{0.107} & \textbf{0.062} & 0.015 \\
 &DQ &11.782 & 13.015 & 7.838 & 8.558 & 10.076 & 3.421 & & 10.578 & 14.582 & 8.282 & 8.101 & 10.076 & 4.249 \\
 &$p$-val &0.067 & 0.043 & 0.250 & 0.200 & 0.121 & 0.755 & & 0.102 & 0.024 & 0.218 & 0.231 & 0.121 & 0.643 \\
 &\\
&$\mathbf{RGumb_{GAS}}$ \\
 &$\hat{C}_{\alpha}$ &0.016 & 0.064 & 0.117 & 0.890 & 0.934 & 0.990 & & 0.016 & 0.062 & 0.112 & 0.890 & 0.938 & 0.990 \\
 &$\hat{L}$ &\textbf{0.025} & \textbf{0.082} & 0.131 & \textbf{0.106} & \textbf{0.060} & 0.015 & & 0.024 & 0.081 & 0.131 & \textbf{0.106} & \textbf{0.060} & 0.015 \\
 &DQ &5.654 & 4.939 & 5.976 & 12.658 & 9.639 & 3.969 & & 5.654 & 3.555 & 5.374 & 12.658 & 5.306 & 3.969 \\
 &$p$-val &0.463 & 0.552 & 0.426 & 0.049 & 0.141 & 0.681 & & 0.463 & 0.737 & 0.497 & 0.049 & 0.505 & 0.681 \\
 &\\
&$\mathbf{N_{GAS}}$ \\
 &$\hat{C}_{\alpha}$ &0.018 & 0.062 & 0.115 & 0.893 & 0.944 & 0.990 & & 0.018 & 0.061 & 0.115 & 0.895 & 0.947 & 0.992 \\
 &$\hat{L}$ &0.025 & \textbf{0.081} & \textbf{0.131} & 0.105 & \textbf{0.059} & 0.015 & & 0.025 & 0.081 & 0.131 & \textbf{0.106} & 0.059 & 0.015 \\
 &DQ &6.695 & 2.735 & 5.886 & 14.911 & 5.558 & 3.969 & & 6.695 & 2.640 & 5.623 & 15.042 & 4.470 & 0.617 \\
 &$p$-val &0.350 & 0.841 & 0.436 & 0.021 & 0.474 & 0.681 & & 0.350 & 0.853 & 0.467 & 0.020 & 0.613 & 0.996 \\
 &\\
&$\mathbf{t_{GAS}}$ \\
 &$\hat{C}_{\alpha}$ &0.016 & 0.066 & 0.113 & 0.893 & 0.946 & 0.993 & & 0.015 & 0.061 & 0.112 & 0.895 & 0.949 & 0.997 \\
 &$\hat{L}$ &0.025 & 0.081 & 0.131 & 0.106 & 0.060 & 0.015 & & 0.024 & 0.081 & 0.131 & 0.105 & 0.059 & 0.015 \\
 &DQ &5.654 & 3.298 & 3.670 & 14.911 & 4.723 & 0.450 & & 4.906 & 2.336 & 3.343 & 14.616 & 4.060 & 3.582 \\
 &$p$-val &0.463 & 0.771 & 0.721 & 0.021 & 0.580 & 0.998 & & 0.556 & 0.886 & 0.765 & 0.023 & 0.669 & 0.733 \\

 \bottomrule
\end{tabular*}
\label{tab1VaR}
\end{table}

Quantile forecasts are central to risk management decisions due to a widespread Value-at-risk (VaR) measurement. VaR is defined as the maximum expected loss that may be incurred by a portfolio over some horizon with a given probability. Let $q_t^{\alpha}$ denote a $\alpha$ quantile of a distribution. VaR of a given portfolio at time $t$ is then simply 
\begin{equation}
q_t^{\alpha} \equiv  F_t^{-1} ({\alpha}), \text{for } {\alpha} \in (0,1).
\end{equation}
Thus, the choice of the distribution is crucial to the VaR calculation. For example, assuming normal distribution may lead to underestimation of the VaR. Our objective is to estimate one-day-ahead\footnote{It is possible to estimate $h$-step-ahead forecasts as well, but these are interesting when a rolling scheme is utilized for forecasting. As explained previously, due to the computational burden of the estimation methodology, we employ static forecasts. In addition, $h$-step ahead forecast requires simulation of the conditional distribution from the model; hence, computational intensity would increase with the horizon employed.} VaR of an equally weighted portfolio composed from oil and stock returns $Y_t = 0.5 X_{1t} + 0.5 X_{2t}$, which have conditional time-varying joint distribution $F_t$. In the previous analysis, we found that the realized GARCH model with time-varying GAS copulas well fits and forecasts the data; thus, we utilize it in VaR forecasts to see if it also correctly forecasts the joint distribution. As there is no analytical formula that can be utilized, the future conditional joint distribution is simulated from the estimated models. Once we obtain the future distribution of the portfolio, the VaR is computed from the corresponding quantile.

While quantile forecasts can be readily evaluated by comparing their actual (estimated) coverage 
$\hat{C}_{\alpha} = 1/n \sum_{n=1}^T 1(y_{t,t+1}<\hat{q}^{\alpha}_{t,t+1})$, against their nominal coverage rate, 
$C_{\alpha} = E[1(y_{t,t+1}<q^{\alpha}_{t,t+1})]$, this approach is unconditional and does not capture the possible dependence in the coverage rates. The number of approaches has been proposed for testing the appropriateness of quantiles conditionally; for the best discussion, see \cite{berkowitz2011evaluating}. In our out-of-sample VaR testing, we employ an approach originally proposed by \cite{engle2004caviar}, who use the $n$-th order autoregression
\begin{equation}
I_{t} = \omega + \sum_{k=1}^n \beta_{1k}I_{t-k} + \sum_{k=1}^n \beta_{2k} q^{\alpha}_{t-k+1} + u_t,
\end{equation}
where $I_{t+1}$ is $1$ if $y_{t+1}<q_t^{\alpha}$, and zero otherwise. While the hit sequence $I_t$ is a binary sequence, $u_t$ is assumed to follow a logistic distribution, and we can estimate it as a simple logit model and test whether $Pr(I_t = 1) = q_t^{\alpha}$. To obtain the $p$-values, we rely on simulations as suggested by \cite{berkowitz2011evaluating}, and we refer to this test as a $DQ$ test in the results. 

The main motivation of the DQ test is to determine whether the conditional quantiles are correctly dynamically specified; hence, it evaluates the absolute performance of the various models.  To assess the relative performance of the models, we evaluate the accuracy of the VaR forecasts statistically by defining the expected loss of the VaR forecast made by a forecaster $m$ as
\begin{equation}
L_{\alpha,m} = E\left[ \alpha - 1\left( y_{t,t+1}<q^{\alpha,m}_{t,t+1} \right) \right] \left[ y_{t,t+1}-q^{\alpha,m}_{t,t+1} \right],
\end{equation},
which was proposed by \cite{giacomini2005evaluation}. The tick loss function penalizes quantile violations more heavily, and the penalization increases with the magnitude of the violation. As argued by \cite{giacomini2005evaluation}, the tick loss is a natural loss function when evaluating conditional quantile forecasts. To compare the forecast accuracy of the two models, we test the null hypothesis that the expected losses for the models are equal, $H_0: d=L_{\alpha,1}-L_{\alpha,2}=0$, against a general alternative. The differences can be tested using \cite{diebold2002comparing} test statistics, $S=\overline{d}/\sqrt{\widehat{LRV}/T}$, where $\overline{d}$ is the unconditional average of loss difference $d$, and $\widehat{LRV}$ a consistent estimate of the long-run variance of $\sqrt{T}\overline{d}$. Under the null of equal predictive accuracy, $S\sim N(0,1)$

Table \ref{tab1VaR} reports the out-of-sample VaR evaluation of all models. As standard normal distribution is the most common choice for VaR computations, we also report the results for the constant normal copula. In addition, we benchmark all the models to versions with the GJR GARCH, which does not employ high frequency data. We can see that all the time-varying models are well specified, and the conditional quantile forecasts from them are not rejected by the DQ test. This holds for both the GJR GARCH as well as the realized GARCH in margins. With the constant copula model, quantile forecasts are rejected primarily for the lower quantiles, and according to the empirical conditional rates, we can see that it underestimates the risk. 

For statistical testing, we employ time-varying Student's $t$ as a benchmark forecaster and test all the other models against it. When looking at the loss functions $\hat{L}_{\alpha,m}$, we can see that the constant copula model is usually rejected against the time-varying Student's $t$. This is also the case for other two time-varying specifications, and so, the realized GARCH time-varying Student's $t$ copula model appears to have statistically the most accurate quantile forecasts. Interestingly, when looking at the results from semiparametric models, we see fewer rejections, and overall, these models appear to provide few more accurate quantile forecasts. The results remain almost the same when the model is benchmarked to the GJR GARCH specifications. The GJR GARCH overestimates the VaR at all quantiles, while the use of high frequency data help at the right tail of the distribution, where the realized GARCH models outperform the GJR GARCH models. While one would expect high frequency data to improve the forecasts, we note that this may be a feature of the static nature of the forecasts. Even in a static environment, a high frequency measure statistically outperforms the benchmark, and while we are evaluating the VaR forecasts, this result has direct economic implications for the improvement of dynamic hedging.

\subsection{Time varying diversification benefits}

In case the dependence of the assets is strongly changing over time, it needs to translate into the changing of diversification benefits as well. While mean dependence is employed in most of the studies to assess the diversification benefits, independence in tails may translate into higher than anticipated benefits. In case the empirical distribution departs from normality, it is important to also account for this departure when calculating diversification benefits. In the previous section, we have observed that the empirical model we have built captures the quantiles of the return distribution well and is correctly specified. The correct choice of the model for quantiles is also important for identification of diversification benefits, which we will utilize here. 

Unlike VaR, the expected shortfall satisfies the sub-additivity property and is a coherent measure of risk. Motivated by these properties, \cite{christoffersen2012potential} propose a measure capturing the dynamics in diversification benefits based on expected shortfall. The conditional diversification benefit (CDB) for a given probability level $\alpha$ is defined by
\begin{equation}
CDB_t^{\alpha} = \frac{\overline{ES}_t^{\alpha}-ES_t^{\alpha}}{\overline{ES}_t^{\alpha}-\underline{ES}_t^{\alpha}},
\end{equation}
where $ES_t^{\alpha}$ is the expected shortfall of the portfolio at hand,
\begin{equation}
ES_t^{\alpha} \equiv E[Y_t|F_{t-1}, Y_t \leq F_t^{-1} ({\alpha})], \text{for } \alpha \in (0,1),
\end{equation}
$\overline{ES}_t^{\alpha}$ is the upper bound of the portfolio, the expected shortfall being the weighted average of the asset's individual expected shortfalls, and $\underline{ES}_t^{\alpha}$ the lower bound on the expected shortfall being the inverse cumulative distribution function for the portfolio. In other words, this lower bound corresponds to the case where the portfolio never loses more than its $\alpha$ distribution quantile. The measure is designed to stay within $[0,1]$ interval and is increasing in the level of diversification benefits. When the CDB is equal to zero, there are literarily no benefits from diversification; when it equals one, the benefits from diversification are the highest possible. 

\begin{figure}[h]
\begin{center}
\includegraphics[width=0.8\textwidth]{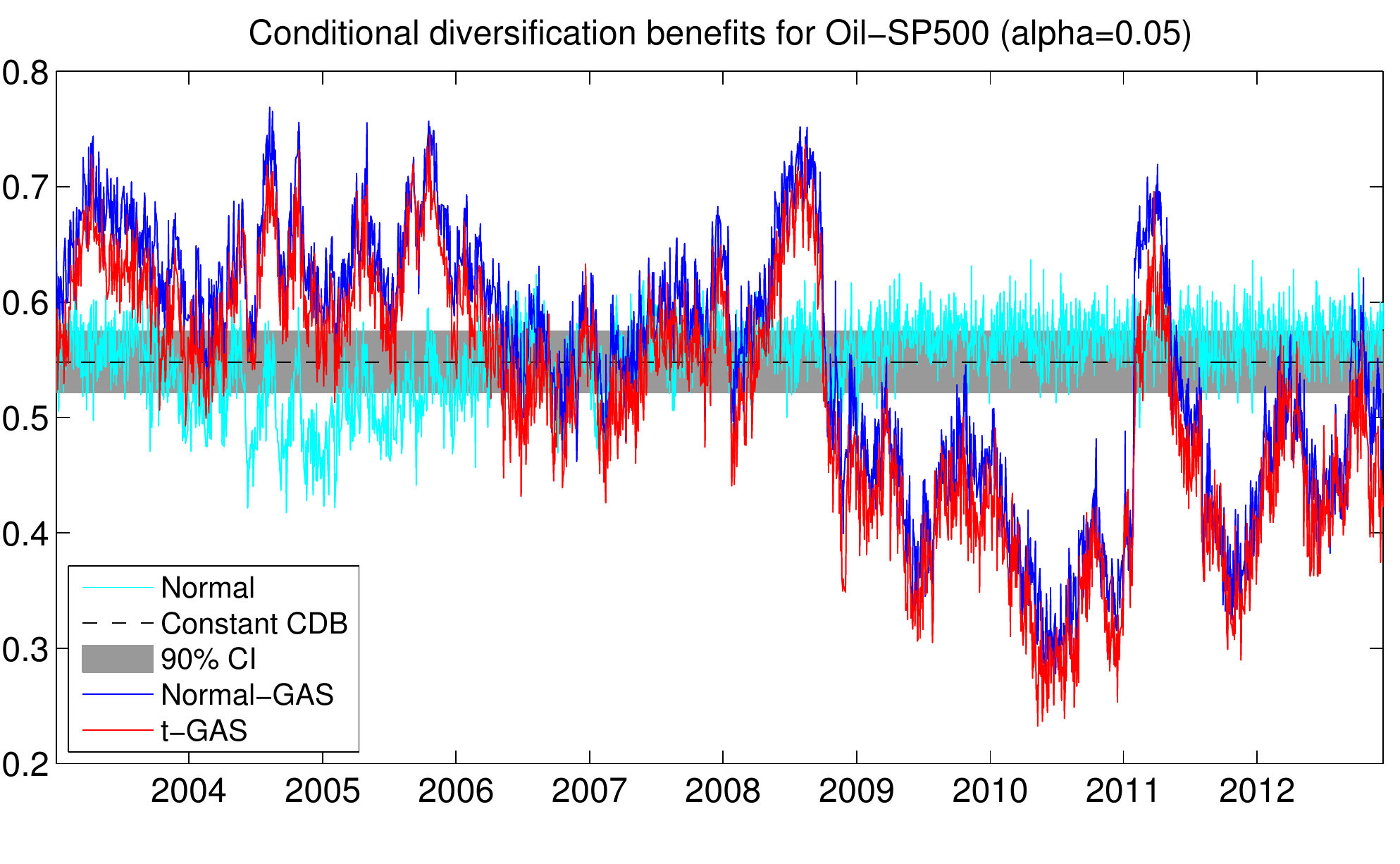}
\end{center}
\caption{Conditional diversification benefits, $CDB_t^{0.05}$ utilizing constant normal, time-varying normal, and StudentÕs $t$ copulas. The dashed line is the constant level of diversification benefits, and in grey, its 90\% confidence band. }
\label{fig:cdb}
\end{figure}

Figure \ref{fig:cdb} plots the conditional diversification benefits for the oil and stocks portfolio implied by the two best performing models in the VaR evaluation for $\alpha = 0.05$. The correct dynamic specification of the quantiles from the empirical models is crucial for the CDB, as it mainly captures the potential diversification benefits from tail independence. Hence, to obtain a precise CDB, one needs to first find a model that captures the dynamics in the quantiles correctly. 

Similarly to the VaR case, as there is no closed form to our empirical model, we need to rely on the simulations for computing the CDB. Encouraged by the previous results, we compute the CDB for the best performing models with time-varying normal and student's $t$ copulas. As a benchmark, we include the model with a constant copula. Moreover, we report 90\% bootstrapped confidence bands computed around a constant level of diversification benefits. Assuming the returns data are independently distributed over time with the same unconditional correlation as the oil and stocks pair, the bootstrap confidence level can be conveniently computed with simulations. We use 10,000 simulations and report the mean value together with the distribution of the constant conditional benefits.

From the Figure \ref{fig:cdb}, we can see how greatly the diversification benefits vary over time. Corresponding to the correlations, there are also several identifiable periods where the benefits from diversification were significantly different. Thus, we conduct the same endogeneity test to find whether there is a structural break in the CDS, and the result is that the test identified exactly the same dates as using the correlation. In addition, January 31, 2011 was identified as another structural break. 

While in the first period the benefits from diversification were relatively high, in the second period between 2006 and 2008, they decreased corresponding to increasing correlation. In the several years after the 2008 crisis, benefits from diversification between oil and stocks were decreasing rapidly, while we can see some rebound in the last few years.

\section{Conclusions}
This work revisits the oil and stocks dependence with the aim of studying the opportunities of these two assets in portfolio management. We propose utilizing the high frequency data in the copula models by choosing to model the marginal dependence with the realized GARCH of \cite{Hansen2012}. Based on the recently proposed generalized autoregressive score copula functions \citep{Creal2013}, we build a new empirical model for oil and stocks, the realized GARCH time-varying GAS copula.

The modeling strategy is able to capture the time-varying conditional distribution of the oil stocks pair accurately, including the dynamics in the correlation and tails. This also translates into accurate quantile forecasts from the model, which are central to risk management, as they represent value-at-risk. Utilizing the ten years of the data covering several different periods, we study the time-varying correlations, and we find two main endogenous breaks in the dependence structure. Most important, we translate the results into the conditional diversification benefits measure recently proposed by \cite{christoffersen2012potential}. The main result is that the possible benefits from using oil as a diversification tool for stocks have been decreasing rapidly over time, while in the last year of the sample, it displayed some rebound. These results have important implications for the risk industry and portfolio management as commodities have recently become an attractive opportunity for risk diversification in portfolios. According to our results, the benefits may not be as high as in the first half of the sample.

In conclusion, during the period under research, oil and stocks could be used in a well-diversified portfolio less often than common perception would imply. We find substantial evidence of dynamics in tail dependence, which translates into dynamically decreasing diversification benefits from employing oil as a hedging tool for stocks. The empirical results have important implications for portfolio management, which should be explored in the future. It may be useful to think about the improvement in dynamic hedging strategies, which will account for our empirical findings. An interesting venue of research is the inclusion of a threshold or multiple component dependences in the models.

\appendix
\section{Copula functions}
\label{appA}
\subsection{Normal copula}
The Normal copula does not have a simple closed form. For the bivariate case and $|\rho|<1$, we can approximate it by the double integral:
\begin{align}
	C^{N}_{\rho}(u,v)=&\int_{-\infty}^{\Phi^{-1}(u)} \int_{-\infty}^{\Phi^{-1}(v)} \frac{1}{2\pi \sqrt{(1-\rho^2)}} exp \left\{ \frac{-(r^2-2\rho r s +s^2)}{2(1-\rho^2)}\right \} dr ds \\
	& -1<\rho<1 \nonumber
	\end{align}
	where $\Phi^{-1}$ is the inverse of standard normal distribution (\emph{c.d.f}). The correlation is modeled by the transformed variable $\rho_t = (1-e^{-\kappa_t})(1+e^{-\kappa_t})^{-1}$, which guarantees that $\rho_t \ will remain in (-1,1)$. For $\rho=0$, we obtain the independence copula, and for $\rho=1$, the comonotonicity copula. For $\rho=-1$ the countermonotonicity copula is obtained. We note that Normal copula has no tail dependence for $\rho<1$.
\subsection{Student's $t$ copula}
The bivariate Student's $t$ copula is defined by
\begin{equation}
C_{\eta,\rho}^{t}(u,v)=\int_{-\infty}^{t_{\eta}^{-1}(u)} \int_{-\infty}^{t_{\eta}^{-1}(v)} \frac{1}{2\pi \sqrt{1-\rho^2}}  \left(1+ \frac{r^2-2\rho r s+s^2}{\eta(1-\rho^2)} \right)^{-\frac{\eta+2}{2}}dr ds
\end{equation}
where $\rho \in (-1,1)$, and $0<\eta$. The $t_{\eta}^{-1}$ is the inverse of \emph{t} distribution with $\eta$ degrees of freedom. The correlation parameter of the \emph{t} copula undergoes the same transformation as in the case of the Normal to guarantee $\rho_t \in (-1,1)$. For the time varying \emph{t} copula, we allow only the correlation to vary through time, the degrees of freedom $\eta$ remain constant.

In contrast to the Normal copula, provided that $\rho>-1$, the \emph{t} copula has symmetric tail dependence given by
\begin{equation}
\lambda^L=\lambda^U=2t_{\eta+1}\left(-\sqrt{\frac{(\eta+1)(1-\rho)}{1+\rho}}\right)
\end{equation}
We utilize the time-varying dynamics of the correlation $\rho_t$ for the time-varying tail dependence $\lambda_t$.

\subsection{Clayton copula}
The bivariate Clayton copula is defined as 
\begin{equation}
C_\theta^{Cl}(u,v)=(u^{-\theta}+v^{-\theta}-1)^{\frac{-1}{\theta}}, \qquad 0<\theta<\infty 
\end{equation}
In the limit as $\theta \rightarrow 0$, we approach the independence copula, and as $\theta \rightarrow \infty$, we approach the two-dimensional comonotonicity copula.

\subsection{(Rotated) Gumbel copula}
The Gumbel copula is defined by
\begin{equation}
C_\delta^{Gu}(u,v)= \exp\{-((-\log u)^\delta + (- \log v)^\delta)^{1/\delta}\},  \quad 1 \leq \delta < \infty
\end{equation}
The Gumbel copula parameter is required to be greater than one, and the transformation $\delta_t=1+\exp(\kappa_t)$ guarantees this. For $\delta=1$. The Gumbel copula reduces to the fundamental independence copula:
\begin{align*}
C_\delta^{Gu}(u,v)&= exp\{-((-\log u)^1 + (- \log v)^1)^{1/1}\}\\
			&=exp\{ \log u + \log v\} \\
			&=exp\{ \log (uv)\}=uv
\end{align*}
The rotated Gumbel copula has the same functional form as the Gumbel copula and is obtained by replacing \emph{u} and \emph{v} by \emph{1-u} and \emph{1-v}, respectively.

\subsection{Symmetrized Joe-Clayton Copula}
The SJC copula is obtained from the linear combination of the Joe-Clayton copula ($C^{JC}$).
	\begin{equation*}	
	C^{SJC}(u,v|\tau^U,\tau^L)=0.5\cdot(C^{JC}(u,v|\tau^U,\tau^L)+C^{JC}(1-u,1-v|\tau^L,\tau^U)+u+v-1)
	\end{equation*}
	where
\begin{align}
	C^{JC}(u,v|\tau^U,\tau^L)=& 1-(1-\{[1-(1-u)^\psi]^{-\gamma}+[1-(1-v)^\psi]^{-\gamma}-1\}^{-1/\gamma})^{1/\psi} \label{eq:jc}\\
	\psi=&1/\log_2(2-\tau^U) \nonumber\\ 
	\gamma=&-1/\log_2(\tau^L)\nonumber\\ 
	\tau^U\in&(0,1), \quad \tau^L \in (0,1) \nonumber
	\end{align}
	This copula has two parameters, $\tau^U$ and $\tau^L$, representing the upper and lower tail dependence, respectively. For more details on this copula, see \cite{Patton2006}. 	

\clearpage
\section{Figures and Tables}
\label{appB}
\begin{figure}[h]
\begin{center}
\includegraphics[width=0.8\textwidth]{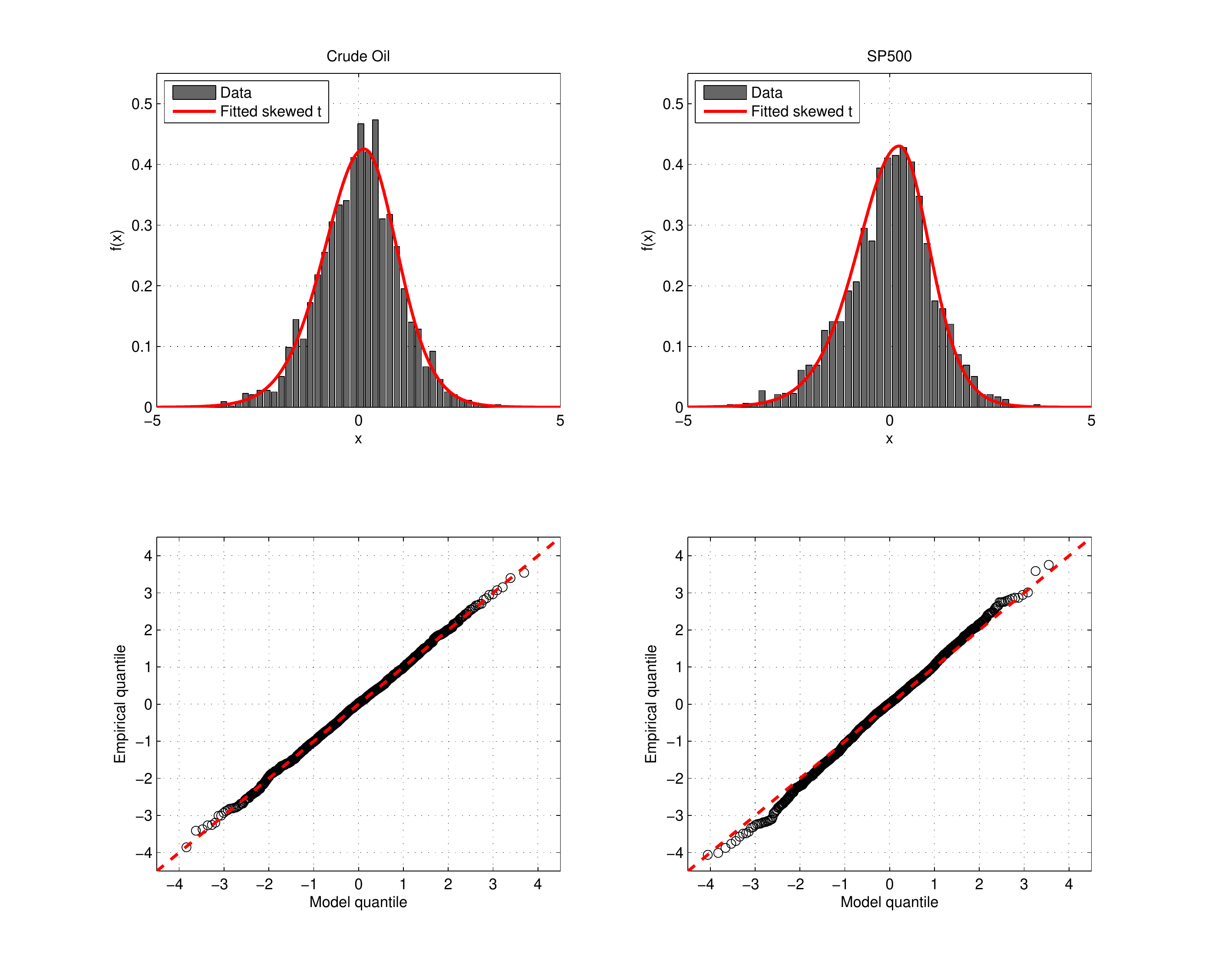}
\end{center}
\label{fig:fittedMargins}
\caption{First row shows fitted skew-\emph{t} density and the histogram of standardized residuals for Crude Oil and the S\&P 500. In the second row, \emph{QQ} plot is shown.}
\end{figure}
\begin{table}[h] 
\footnotesize
\begin{tabular*}{\textwidth}{@{\extracolsep{\fill}}lrrrrrr}
\toprule 
           &  \multicolumn{2}{c}{Returns}  &   \multicolumn{2}{c}{Realized Volatility} \\ 
           \cmidrule{2-3}  \cmidrule{4-5}
           &  Crude Oil  &      S\&P 500  &   Crude Oil  &     S\&P 500  \\ 
           \cmidrule{2-3}  \cmidrule{4-5}
Mean        &      0.000  &      0.000 		 & 0.016 & 0.007\\
Std dev     &      0.017  &      0.010  		& 0.007 & 	0.006 \\
Skewness   &     -0.083  &     -0.352 		& 2.395 &	3.556 \\
Ex. Kurtosis    &      3.924  &     10.940 	& 8.049 &	19.944 \\
Minimum &  -0.108 	&	-0.082		& 0.005 &	0.001 \\
Maximum & 0.123	& 	0.073	 & 0.064 &	0.076 \\
  \bottomrule  
\end{tabular*}   
\caption{Descriptive Statistics for daily oil and stock (S\&P 500) returns and realized volatilities ($\sqrt{RV_t}$) over the sample period extending from January 3, 2003 until December 11, 2012.}
\label{tab:sumStat} 
\end{table} 

\newpage
{\footnotesize{
\setlength{\bibsep}{3pt}
\bibliographystyle{chicago}
\bibliography{Bibliography}
}}

\end{document}